\documentclass[11pt,letterpaper]{article}

\usepackage{amsmath,amssymb,amsthm}
\usepackage[dvipsnames]{xcolor}
\usepackage{algorithm}
\usepackage{algpseudocode}
\usepackage{tikz}
\usetikzlibrary{arrows.meta,positioning,fit,shapes.geometric}
\usepackage[margin=1in]{geometry}
\usepackage[authoryear,square,semicolon]{natbib}
\usepackage{tcolorbox}
\usepackage[colorlinks=true,
    linkcolor=PineGreen,
    citecolor=RedViolet,
    urlcolor=PineGreen]{hyperref}
\theoremstyle{definition}
\newtheorem{theorem}{Theorem}
\newtheorem{lemma}[theorem]{Lemma}
\newtheorem{corollary}[theorem]{Corollary}
\title{Computing Normal-Form Correlated Equilibria in Extensive-Form Games is in PPAD}

\date{}

\author{
  Vincent Cheval\\
  University of Oxford, UK
  \and
  Florian Horn\\
  CNRS, IRIF, France
  \and
  Soumyajit Paul\\
  University of Oxford, UK
  \and
  Mahsa Shirmohammadi\\
CNRS, IRIF, France
}

\begin{document}

\maketitle
\begin{abstract}
We study the computational complexity of 
finding any normal-form correlated equilibria (NFCE for short) in the succinct games of  extensive-form with perfect recall, which is considered as a major open problem in algorithmic game theory~\citep{anagnostides2026complexity,daskalakis2025efficient,peng2024fast,farina2023linear,huang2008computing,papadimitriou2008computing}.
Our main results place this problem in PPAD for an arbitrary numbers of players.
Our second result establishes PPAD-hardness of computing such an NFCE with  support size of at most  quadratic in the size of the game, even for two-player games.
The posed restriction on the support size in PPAD-hardness construction is motivated by our PPAD algorithm, which constructs an NFCE with support size quadratic in the size of the input extensive-form game.
\end{abstract}

\paragraph{AI disclosure.} GPT Astra 6.2 was used interactively to explore and derive the results presented in this paper. GPT Astra 6 was used to improve the clarity and presentation. The authors take full responsibility for the content of the paper.

{\footnotesize 

\tableofcontents
}

\section{Introduction}

Nash equilibrium and correlated equilibrium are two fundamental solution concepts in game theory.
While Nash equilibrium is about independent strategic choices, correlated equilibrium, introduced by Aumann \citeyearpar{aumann1974subjectivity}, allows players to coordinate through a mediator.
The mediator samples a joint strategy profile from a publicly known distribution and privately communicates each player's strategy.
A correlated equilibrium requires that no player benefit from deviating from their recommendation, given that the other players follow theirs.
Two basic computational questions arise~\citep{papadimitriou2008computing,cheval2026complexity}:
\begin{itemize}
    \item the \textsc{Any NFCE} problem, which asks to find an equilibrium, and
    \item the \textsc{Threshold NFCE} problem, which asks whether an equilibrium with social welfare greater than a given threshold exists.
\end{itemize}

For explicitly represented normal-form games, the correlated-equilibrium problem is solvable in polynomial time by linear programming, in contrast to the PPAD-completeness of finding a Nash equilibrium in bimatrix games~\citep{chen2009settling}. 
However,  extensive-form games give a more succinct representation of sequential interactions, where the normal form of the game may be exponentially larger than the input tree. 
In our recent work~\citep{cheval2026complexity}, we extensively studied various notions of correlated equilibria in extensive-form games. 
Among these, the normal-form correlated equilibrium (NFCE) is the direct extension of correlated equilibrium to this setting, with entire strategies recommended before play. 
We resolved several gaps in the complexity landscape of the \textsc{Threshold} problem, establishing PSPACE-hardness for NFCE and NP-completeness for several other correlated-equilibrium notions. 
This lower bound for NFCE was further strengthened to EXPTIME-hardness in a subsequent work~\citep{horn2026threshold}.

Despite these advances, the exact complexity of the \textsc{Any NFCE} problem has remained a major open probelm~\citep{papadimitriou2008computing,vonstengel2008extensive,huang2008computing,farina2023linear,peng2024fast,daskalakis2025efficient,cheval2026complexity}. 
In this paper, we investigate this remaining problem for perfect-recall extensive-form games, both without a restriction on the output support and with prescribed quadratic support size bounds.
Our main results place the \textsc{Any NFCE} problem for these games in PPAD for an arbitrary numbers of players.
Our second result establishes PPAD-hardness of computing such an NFCE with  support size of at most  quadratic in the size of the game, even for two-player games.
The posed restriction on the support size in PPAD-hardness construction is motivated by our PPAD algorithm, which constructs an NFCE with support size quadratic in the size of the input normal-form game\footnote{This draft  is a  work in progress.  We will  further refine the presentation and exposition in subsequent versions of the manuscript.}.

\subsection*{Related Work}
\label{subsec:related-work}

\paragraph{Computation of Correlated equilibria}
The computation of correlated equilibria in succinctly represented
games was initiated by \citet{papadimitriou2008computing}, with
subsequent developments on exact equilibrium computation
\citep{papadimitriou2010comment,jiang2015polynomial}.
In extensive-form games, the exponentially large pure-strategy
spaces have motivated alternative notions of correlation.
\citet{vonstengel2008extensive} introduced extensive-form correlated
equilibrium (EFCE), which admits polynomial-time computation
\citep{huang2008computing,jiang2015polynomial}.
More recently, \citet{daskalakis2025efficient} established efficient
computation of linear correlated equilibria in convex games.
These notions, however, do not capture the full deviations permitted
by NFCE.

\paragraph{Learning correlated equilibria.}
The connection between regret minimization and correlated equilibria
goes back to \citet{hart2000adaptive} and the reduction from external
to swap regret of \citet{blum2007external}.
While linear-swap regret can be minimized efficiently in sequential
games \citep{farina2023linear}, recent advances in swap-regret
minimization \citep{dagan2024external,peng2024fast} yield polynomial-time
approximation schemes for NFCE in extensive-form games.
However, their dependence on the approximation parameter precludes
their direct use for exact equilibrium computation.
Recent work also establishes lower bounds for swap-regret minimization
in tree-form decision problems \citep{anagnostides2026complexity}
and efficient NFCE approximation in finite-horizon Markov games
with a fixed number of players \citep{anagnostides2026markov}.

\paragraph{PPAD and Equilibrium computation.}
Following the introduction of PPAD by
\citet{papadimitriou1994complexity}, computing Nash equilibria,
even approximately, has been shown to be PPAD-complete
\citep{chen2009settling}.
For two-player extensive-form games with perfect recall, sequence-form
methods enable exact Nash equilibrium computation within PPAD
\citep{kollermegiddo1992complexity,kollervonstengel1996efficient}.
For multiplayer games, \citet[Theorem~19, Part~2]{etessami2021complexity}
establishes PPAD membership for computing an almost subgame-perfect
equilibrium, and hence a weak approximate Nash equilibrium, with
binary-encoded accuracy.
Our upper bound builds on this result and the efficient construction
of small-support realization-equivalent mixed strategies
\citep{kollermegiddo1996small} to obtain an exact NFCE.

\paragraph{Support bounds and hardness.}
Small-support correlated equilibria have been studied in normal-form
games \citep{germanolugosi2007,babichenko2014small,soleymani2026support}
and separable games \citep{stein2007thesis}.
In particular, Stein's rank-based bound implies the existence of
quadratic-support NFCEs in extensive-form games.
Our contribution is to establish their computability within PPAD.
While minimum-support correlated equilibrium computation is NP-hard
under Cook reductions \citep{babichenko2014small}, hardness results
for sparse coarse correlated equilibria
\citep{foster2023hardness,anagnostides2024sparse}
concern mixtures of a small number of product distributions.
More closely related to our work,
\citet{anagnostides2026complexity} establish PPAD-hardness of
approximate correlated equilibrium computation under prescribed
support bounds in multilinear games on rational polytopes.
Our lower bound establishes PPAD-completeness for exact NFCE
under prescribed support bounds in explicitly represented
two-player extensive-form games.

\subsection{Overview of Main  Results}
Let $V$ be the set of nodes in the game tree, $N$ the number of terminal nodes, and $n$ the number of players. Write
\[
d_i:=\sum_{I\in\mathcal I_i} (|A(I)|-1),\qquad
K:=1+\sum_{i=1}^n d_i(d_i+1)\le 1+N(N-1).
\]
Here $d_i$ is the affine dimension of the realization polytope of player $i$ (Section~\ref{sec:overview-ppad}), $\mathcal I_i$ is the set of its information sets and $A(I)$ is the set of actions available at $I$. All outputs are explicit families of full pure-plan profiles with rational probabilities of polynomial encoding length.

\begin{theorem}\label{thm:main-upper}
Computing an  NFCE of a rational extensive-form game with perfect recall lie in  $\mathrm{PPAD}$, even when the number of players is part of the input. The reduction outputs a (rational) NFCE with support size at most~$K\le 1+N(N-1)$.
\end{theorem}

\begin{corollary}\label{cor:existence}
Every game in this setting admits an exact rational NFCE of polynomial encoding length satisfying the support bound of Theorem~\ref{thm:main-upper}.
\end{corollary}

For the hardness result, we use the support size bounds
\[
\kappa_d:=(d_1+1)(d_2+1),\qquad
\kappa_V:=\left\lfloor\frac{|V|^2}{36}\right\rfloor,
\qquad
\kappa_+:=\max\{\kappa_d,\kappa_V\}.
\]
\begin{theorem}\label{thm:main-hardness}
For two-player perfect-recall extensive-form games, computing an  NFCE under either support size bound $\kappa_d$ or $\kappa_+$ is $\mathrm{PPAD}$-complete.
\end{theorem}

The  support size bound $\kappa_V$ alone need not guarantee existence on arbitrary inputs; the maximum in $\kappa_+$ ensures totality. Theorem~\ref{thm:main-hardness} concerns these specified support size bounds, not arbitrary constant multiples of $|V|^2$ or arbitrary polynomial bounds. 
\section{Preliminaries}
\label{sec:preliminaries}

\paragraph{Extensive-form games.}
We consider finite, multiplayer stochastic extensive-form games
with imperfect information and perfect recall.
A game $\Gamma$ consists of a finite rooted directed tree
with node set $V$ and terminal nodes $Z\subseteq V$.
Each nonterminal node is controlled either by chance or by
one of the $n$ players, where $n$ is part of the input.
At chance nodes, outgoing edges are assigned probabilities;
each terminal node $z\in Z$ has a payoff vector
$u(z)=(u_1(z),\ldots,u_n(z))\in\mathbb{Q}^n$.
We assume that the tree is explicitly represented and that
all probabilities and payoffs are rational numbers encoded
in binary.

For each player $i\in[n]$, the nodes controlled by $i$
are partitioned into information sets $\mathcal I_i$.
All nodes in an information set $I\in\mathcal I_i$
have the same non-empty set of available actions $A(I)$.
For a node $v$, let $h_i(v)$ be the ordered sequence of
information-set/action pairs of player $i$ along the path
from the root to $v$.
A game has \emph{perfect recall} if
\[
h_i(v)=h_i(v')
\qquad
\text{for every }i\in[n],\
I\in\mathcal I_i,\ v,v'\in I.
\]

\paragraph{Strategies and expected payoffs.}
A pure strategy (full pure plan) of player $i$ is a function
$\sigma_i$ assigning an action $\sigma_i(I)\in A(I)$
to every $I\in\mathcal I_i$.
Let $\Sigma_i$ denote the set of such strategies and
$\Sigma=\prod_{i=1}^n\Sigma_i$ the set of pure-strategy profiles; write $\Sigma_{-i}=\prod_{j\ne i}\Sigma_j$.

A mixed strategy is a distribution
$\alpha_i\in\Delta(\Sigma_i)$.
A mixed-strategy profile is
$\boldsymbol{\alpha}=(\alpha_1,\ldots,\alpha_n)$,
whose induced distribution on $\Sigma$ is the product
$\bigotimes_i\alpha_i$.
A behavioral strategy assigns a distribution over $A(I)$
to each $I\in\mathcal I_i$.
By Kuhn's theorem \citeyearpar{kuhn1953extensive}, mixed and
behavioral strategies are realization-equivalent under
perfect recall.

For a terminal node $z$, let $\mathcal I_i(z)$ be the information sets of player $i$ on the root--$z$ path, and let $a_z(I)$ be the action taken at $I$ along that path. Define
\begin{equation}\label{eq:leaf-compat}
\chi_i(\sigma_i,z):=\prod_{I\in\mathcal I_i(z)}\mathbf 1\{\sigma_i(I)=a_z(I)\}.
\end{equation}
Intuitively, $\chi_i(\sigma_i,z)$ indicates whether player $i$'s plan permits the path to $z$ (its own leaf realization weight; see \citep[Definition~2.4 and Lemma~2.5]{kollermegiddo1996small}).
For a mixed strategy $\alpha_i$ and a behavioral strategy $\beta_i$, respectively, put
\[
\chi_i(\alpha_i,z):=\mathbb E_{\sigma_i\sim\alpha_i}[\chi_i(\sigma_i,z)],
\qquad
\chi_i(\beta_i,z):=\prod_{I\in\mathcal I_i(z)}\beta_i(I)(a_z(I)).
\]
We call two strategies of player $i$ \emph{realization-equivalent} when these leaf realization weights coincide for every $z\in Z$. In particular, replacing one strategy by a realization-equivalent strategy preserves the terminal distribution against every fixed opponents' profile.
Let $p_c(z)$ be the product of chance probabilities on that path. The expected payoff for player $i$ of a pure profile $\sigma\in\Sigma$ is
\begin{equation}\label{eq:payoff-leaves}
U_i(\sigma)=\sum_{z\in Z}p_c(z)u_i(z)\prod_{j=1}^n\chi_j(\sigma_j,z).
\end{equation}
We extend $U_i$
%multilinearly
to mixed-strategy profiles and 
%linearly 
to distributions over $\Sigma$.

\paragraph{Nash equilibrium.}
A mixed-strategy profile $\boldsymbol\alpha$ is a \emph{Nash equilibrium} if, for every $i\in[n]$ and $\sigma_i'\in\Sigma_i$,\footnote{It suffices to consider pure deviations by linearity in the deviating player's mixed strategy.}
\[
U_i(\boldsymbol\alpha)\ge U_i(\sigma_i',\boldsymbol\alpha_{-i}).
\]
For $\epsilon\ge0$, it is a \emph{weak $\epsilon$-Nash equilibrium} if
\[
U_i(\sigma_i',\boldsymbol\alpha_{-i})-U_i(\boldsymbol\alpha)\le\epsilon
\qquad( i\in[n],\ \sigma_i'\in\Sigma_i).
\]

A behavioral profile is a weak $\epsilon$-Nash equilibrium if a realization-equivalent mixed profile is one. By perfect recall and the leaf payoff formula, this definition does not depend on the choice of the equivalent mixed strategies.

\paragraph{Normal-form correlated equilibrium.}
A correlation plan is a distribution $\mu\in\Delta(\Sigma)$. A mediator samples a pure profile and privately recommends its entire pure strategy to each player before play begins.

For player $i$, recommended strategy $\sigma_i\in\Sigma_i$ and alternative strategy $\sigma_i'\in\Sigma_i$, define the \emph{unnormalized deviation gain}
\[
R_{i,\sigma_i,\sigma_i'}(\mu):=
\sum_{\tau\in\Sigma:\,\tau_i=\sigma_i}\mu(\tau)
\bigl[U_i(\sigma_i',\tau_{-i})-U_i(\tau)\bigr].
\]
The distribution $\mu$ is an \emph{exact normal-form correlated equilibrium} (NFCE) if
\[
R_{i,\sigma_i,\sigma_i'}(\mu)\le0
\qquad(i\in[n],\ \sigma_i,\sigma_i'\in\Sigma_i).
\]
For $\epsilon\ge0$, it is an \emph{unnormalized $\epsilon$-NFCE} if
\[
R_{i,\sigma_i,\sigma_i'}(\mu)\le\epsilon
\qquad(i\in[n],\ \sigma_i,\sigma_i'\in\Sigma_i).
\]

\paragraph{Deviation rules.}
Equivalently, $\mu$ is an exact NFCE if, for every player $i$ and every deterministic deviation rule $\phi_i:\Sigma_i\to\Sigma_i$,
\begin{equation}\label{eq:global-nfce}
\mathbb E_{\sigma\sim\mu}\bigl[U_i(\phi_i(\sigma_i),\sigma_{-i})-U_i(\sigma)\bigr]\le0.
\end{equation}
Indeed, the gain of a rule is $\sum_{\sigma_i\in\Sigma_i}R_{i,\sigma_i,\phi_i(\sigma_i)}(\mu)$ (cf.\ \citep{stein2007thesis,babichenko2014small}). Conversely, choosing a rule that changes only one recommendation recovers the corresponding unnormalized deviation inequality at zero error. Randomized rules cannot improve on deterministic ones by linearity. We use this exact deviation-rule formulation in the hardness proof; unnormalized approximate NFCEs are used only as an intermediate step in the PPAD algorithm.

\section{PPAD Algorithm}
\subsection{Overview of the algorithm}
\label{sec:overview-ppad}

In this section, we state the ingredients of the algorithm.
The proofs of the support construction and exact reweighting are given in the subsequent subsections.

\begin{theorem}[Weak approximate Nash in PPAD~[\cite{etessami2021complexity}]
\label{thm:approx-nash-ppad}
Given an extensive-form game with perfect recall and a
rational $\epsilon>0$ encoded in binary,
computing a rational behavioral weak $\epsilon$-Nash equilibrium
belongs to $\mathrm{PPAD}$, even when the number of players is
part of the input (for fewer than three players, add inactive players). The output has size polynomial in the size of the game and $\epsilon$.
\end{theorem}

\paragraph{Small realization-equivalent mixtures.}
By \citet[Theorem~4.2]{kollermegiddo1996small}, a rational behavioral strategy admits a polynomial-time computable realization-equivalent mixed strategy over at most $N$ pure strategies. We use the dimension-based refinement of this standard result.

The small-support existence bound $1+\sum_i d_i(d_i+1)$ is has also been shown using rank-based separable-game result \citep[Theorem~3.3.6]{stein2007thesis}: all payoffs depend on player $i$'s strategy through its $d_i$-dimensional realization polytope. Our algorithm obtains the joint support without enumerating the full product of the marginal supports.

\begin{lemma}
\label{lem:product-nash-nfce}
Let $\boldsymbol\alpha$ be a weak $\epsilon$-Nash equilibrium in mixed strategies, with $\epsilon\ge0$. Then the product distribution $\mu=\bigotimes_i\alpha_i$ is an unnormalized $\epsilon$-NFCE.
\end{lemma}
\begin{proof}
Fix $i$ and put $M_i=\max_{\sigma_i'\in\Sigma_i}U_i(\sigma_i',\boldsymbol\alpha_{-i})$. By weak Nash and linearity,
\[
\sum_{\sigma_i\in\Sigma_i}\alpha_i(\sigma_i)
\bigl[M_i-U_i(\sigma_i,\boldsymbol\alpha_{-i})\bigr]
=M_i-U_i(\boldsymbol\alpha)\le\epsilon.
\]
Every summand is nonnegative by the definition of $M_i$. Independence of the recommendations gives, for any $\sigma_i,\sigma_i'\in\Sigma_i$,
\begin{align*}
R_{i,\sigma_i,\sigma_i'}(\mu)
&=\alpha_i(\sigma_i)\bigl[U_i(\sigma_i',\boldsymbol\alpha_{-i})-U_i(\sigma_i,\boldsymbol\alpha_{-i})\bigr]\\
&\le\alpha_i(\sigma_i)\bigl[M_i-U_i(\sigma_i,\boldsymbol\alpha_{-i})\bigr]\le\epsilon.
\end{align*}
The identity also holds when $\alpha_i(\sigma_i)=0$, since both sides are zero.
\end{proof}

The following theorem allows us to use an approximate equilibrium to identify a support and then find exact probabilities on that support. The required accuracy depends only on the input game and the support bound.

\begin{theorem}
\label{thm:exact-support}
Assume that all terminal payoffs are normalized to $[0,1]$.
Define
\[
N:=|Z|,\qquad
K:=1+\sum_i d_i(d_i+1),\qquad
B:=\prod_{r\in\mathcal R}\operatorname{den}(r),
\]
where $\mathcal R$ contains all rational chance-edge
probabilities and all normalized terminal payoffs, counting each
input occurrence, and $\operatorname{den}(r)$ is the positive
denominator of $r$ in lowest terms. Define
\[
\epsilon_\Gamma:=\frac{1}{2((K+1)B)^{K+1}}.
\]
Every unnormalized $\epsilon_\Gamma$-NFCE $\mu$ supported on at
most $K$ full pure-plan profiles admits an exact NFCE $\mu^*$ with
$\operatorname{supp}(\mu^*)\subseteq\operatorname{supp}(\mu)$.
The proof is given in Section~\ref{sec:choosing-epsilon}.
\end{theorem}

\begin{theorem}
\label{thm:main-two}
Let $\epsilon\geq0$ be rational and let
$\boldsymbol\alpha=(\alpha_i)_{i=1}^n$ be a rational weak
$\epsilon$-Nash mixed-strategy profile. For each $i$, put
$\mathcal A_i:=\operatorname{supp}(\alpha_i)$ and
$m_i:=|\mathcal A_i|$, and assume $m_i\leq d_i+1$. There is a polynomial-time algorithm that computes a
rational unnormalized $\epsilon$-NFCE $\mu_0$ supported on
profiles in $\prod_i\mathcal A_i$ such that
\[
|\operatorname{supp}(\mu_0)|\leq 1+\sum_{i=1}^n m_i d_i\leq K,
\]
and, writing $\mu_{\mathrm{prod}}=\bigotimes_i\alpha_i$,
\[
R_{i,\sigma_i,\sigma_i'}(\mu_0)=R_{i,\sigma_i,\sigma_i'}(\mu_{\mathrm{prod}})
\quad(i\in[n],\ \sigma_i,\sigma_i'\in\Sigma_i).
\] The proof is given in Section~\ref{sec:small-mixed-support}.
\end{theorem}

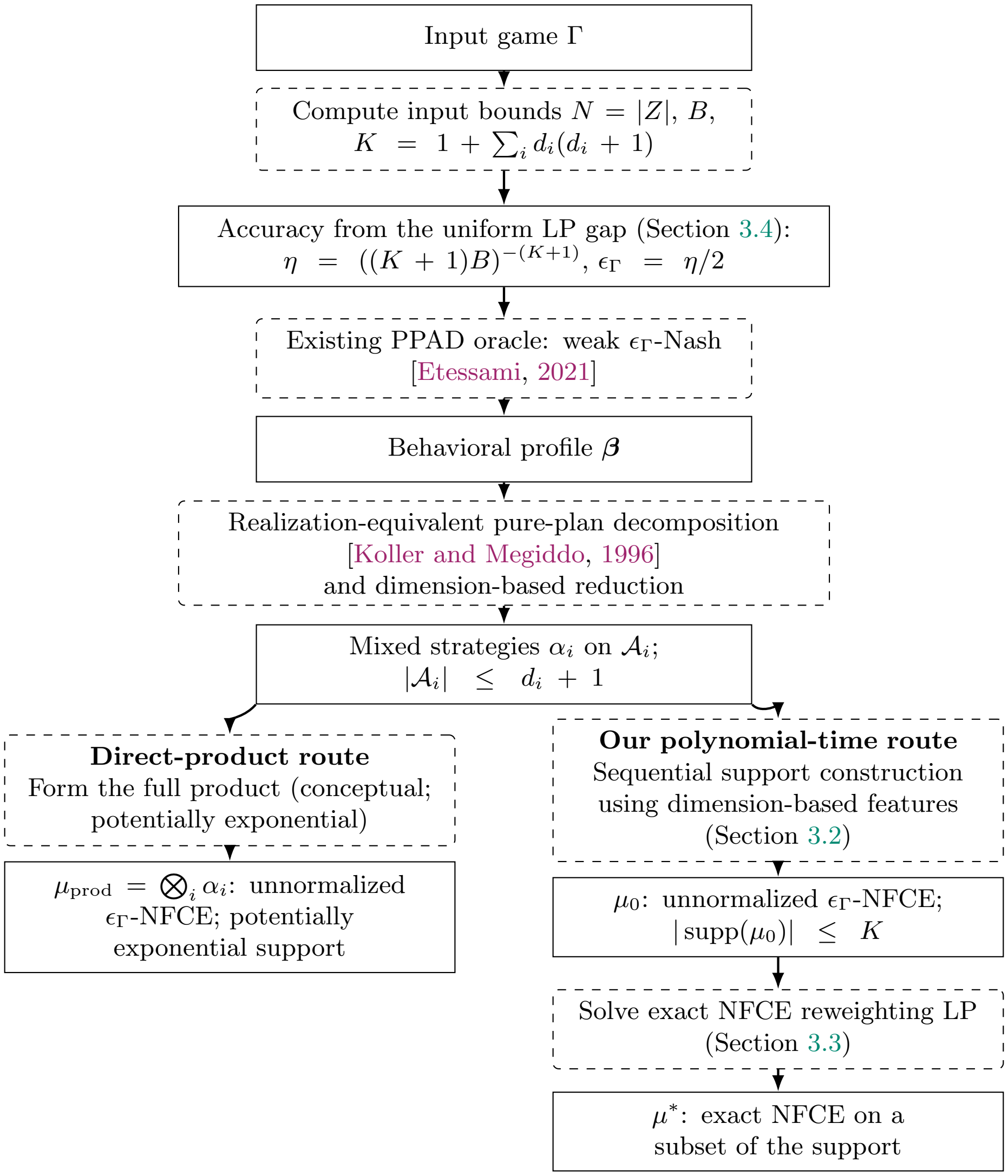
\begin{figure}[htbp]
\centering
\begin{tikzpicture}[
  font=\footnotesize,
  procedure/.style={draw, dashed, rounded corners=2pt, align=center,
    inner xsep=5pt, inner ysep=4pt, minimum height=8mm},
  object/.style={draw, solid, align=center,
    inner xsep=5pt, inner ysep=4pt, minimum height=8mm},
  flow/.style={-{Latex[length=2mm]}, thick},
  pathlabel/.style={font=\footnotesize\bfseries, align=center}
]
  % Common input and established algorithmic ingredients.
  \node[object, text width=5.7cm] (game) at (0,0)
    {Input game $\Gamma$};
  \node[procedure, text width=5.7cm] (bounds) at (0,-1.12)
    {Compute input bounds $N=|Z|$, $B$,\\
     $K=1+\sum_i d_i(d_i+1)$};
  \node[object, text width=7.6cm] (precision) at (0,-2.55)
    {Accuracy from the uniform LP gap (Section~\ref{sec:choosing-epsilon}):\\
     $\eta=((K+1)B)^{-(K+1)}$, $\epsilon_\Gamma=\eta/2$};
  \node[procedure, text width=5.7cm] (oracle) at (0,-3.92)
    {Existing PPAD oracle: weak $\epsilon_\Gamma$-Nash\\
     \citep{etessami2021complexity}};
  \node[object, text width=5.7cm] (behavior) at (0,-5.03)
    {Behavioral profile $\boldsymbol\beta$};
  \node[procedure, text width=7.6cm] (decompose) at (0,-6.30)
    {Realization-equivalent pure-plan decomposition\\
     \citep{kollermegiddo1996small} and dimension-based reduction};
  \node[object, text width=5.7cm] (mix) at (0,-7.66)
    {Mixed strategies $\alpha_i$ on $\mathcal A_i$;\\
     $|\mathcal A_i|\leq d_i+1$};

  \node[procedure, text width=5.15cm] (product) at (-3.35,-9.20)
    {\textbf{Direct-product route}\\
     Form the full product (conceptual;\\
     potentially exponential)};
  \node[object, text width=5.15cm] (epsilonproduct) at (-3.35,-10.75)
    {$\mu_{\mathrm{prod}}=\bigotimes_i\alpha_i$: unnormalized\\
     $\epsilon_\Gamma$-NFCE; potentially exponential support};

  \node[procedure, text width=5.15cm] (compress) at (3.35,-9.20)
    {\textbf{Our polynomial-time route}\\
     Sequential support construction\\
     using dimension-based features\\
     (Section~\ref{sec:small-mixed-support})};
  \node[object, text width=5.15cm] (epsilonsmall) at (3.35,-10.75)
    {$\mu_0$: unnormalized $\epsilon_\Gamma$-NFCE;\\
     $|\operatorname{supp}(\mu_0)|\leq K$};
  \node[procedure, text width=5.15cm] (reweight) at (3.35,-12.12)
    {Solve exact NFCE reweighting LP\\
     (Section~\ref{sec:final-lp})};
  \node[object, text width=5.15cm] (exact) at (3.35,-13.37)
    {$\mu^*$: exact NFCE on a subset of the support};

  \draw[flow] (game) -- (bounds);
  \draw[flow] (bounds) -- (precision);
  \draw[flow] (precision) -- (oracle);
  \draw[flow] (oracle) -- (behavior);
  \draw[flow] (behavior) -- (decompose);
  \draw[flow] (decompose) -- (mix);
  \draw[flow] (mix.south west) to[out=230,in=90] (product.north);
  \draw[flow] (mix.south east) to[out=310,in=90] (compress.north);
  \draw[flow] (product) -- (epsilonproduct);
  \draw[flow] (compress) -- (epsilonsmall);
  \draw[flow] (epsilonsmall) -- (reweight);
  \draw[flow] (reweight) -- (exact);
\end{tikzpicture}
\caption{Overview of the PPAD Algorithm. The left route forms the potentially exponential product distribution, whereas the
right route constructs and exactly reweights a sparse distribution.}
\label{fig:overview}
\end{figure}

\subsection{Small mixed \texorpdfstring{$\epsilon$}{epsilon}-Nash to generate support for NFCE}
\label{sec:small-mixed-support}

We first express deviation gains in terms of leaf contributions, then retain only the coordinates needed to preserve these gains. 

For every $i\in[n]$, $\sigma_i\in\mathcal A_i\subseteq\Sigma_i$, and $z\in Z$, define
\begin{equation}\label{eq:leaf-Phi}
\Phi_{i,\sigma_i,z}(\tau):=\mathbf1\{\tau_i=\sigma_i\}\,p_c(z)u_i(z)\prod_{j\ne i}\chi_j(\tau_j,z).
\end{equation}
Intuitively, $\Phi_{i,\sigma_i,z}$ records the payoff contribution of leaf $z$ when player $i$ receives recommendation $\sigma_i$ and the opponents permit that leaf, before checking player $i$'s own choices.

Let $\mathcal A_i=\operatorname{supp}(\alpha_i)$ and $m_i=|\mathcal A_i|$. The leaf-contribution vector $\Phi(\tau)$ consists of these $|Z|\sum_i m_i$ coordinates.

\begin{lemma}\label{lem:regret-identity}
For any distribution $\mu$ on $\prod_i\mathcal A_i$, every $i\in[n]$, $\sigma_i\in\mathcal A_i$, and every $\sigma_i'\in\Sigma_i$,
\begin{equation}\label{eq:feature-regret}
R_{i,\sigma_i,\sigma_i'}(\mu)=
\sum_{z\in Z}\bigl[\chi_i(\sigma_i',z)-\chi_i(\sigma_i,z)\bigr]\mathbb E_\mu[\Phi_{i,\sigma_i,z}].
\end{equation}
Consequently, preserving the expectation of $\Phi$ preserves all full-plan deviation gains, including deviations outside the mixed supports.
\end{lemma}
\begin{proof}
Substituting the leaf payoff formula \eqref{eq:payoff-leaves} into the definition of $R$ gives
\begin{align*}
R_{i,\sigma_i,\sigma_i'}(\mu)
&=\sum_{\tau:\tau_i=\sigma_i}\mu(\tau)\sum_{z\in Z}
  [\chi_i(\sigma_i',z)-\chi_i(\sigma_i,z)]p_c(z)u_i(z)
  \prod_{j\ne i}\chi_j(\tau_j,z)\\
&=\sum_{z\in Z}[\chi_i(\sigma_i',z)-\chi_i(\sigma_i,z)]
  \sum_{\tau}\mu(\tau)\Phi_{i,\sigma_i,z}(\tau)\\
&=\sum_{z\in Z}[\chi_i(\sigma_i',z)-\chi_i(\sigma_i,z)]\mathbb E_\mu[\Phi_{i,\sigma_i,z}].
\end{align*}
The identity holds for all $\sigma_i'\in\Sigma_i$. If $\sigma_i\notin\mathcal A_i$, its recommendation probability and every associated deviation gain are zero for every distribution on $\prod_i\mathcal A_i$. The consequence follows.
\end{proof}

\paragraph{Dimension-based payoff features.}
The deviation vector $\chi_i(\sigma_i',\cdot)-\chi_i(\sigma_i,\cdot)$ in \eqref{eq:feature-regret} belongs to the direction space of the leaf realization polytope. To compute this space we use results from \cite{kollermegiddo1996small} and map their solution space to the leaf coordinates using the last own pair of each leaf (zero when there is none). The affine-hull argument in Section~\ref{sec:overview-ppad} shows that this image is exactly the span of the differences of pure leaf realization vectors. Rational Gaussian elimination yields a rational basis $b_{i,1},\ldots,b_{i,q_i}$ with $q_i\le d_i$ in polynomial bit time. This uses only the explicit sequence-form equations, not an enumeration of pure plans.
Define the projected feature vector $\Psi(\tau)$ by
\begin{equation}\label{eq:projected-features}
\Psi_{i,\sigma_i,\ell}(\tau):=\sum_{z\in Z}b_{i,\ell}(z)\Phi_{i,\sigma_i,z}(\tau),
\qquad 1\le\ell\le q_i,\quad \sigma_i\in\mathcal A_i.
\end{equation}
The number of coordinates is
\begin{equation}\label{eq:D-def}
D=\sum_i m_iq_i\le\sum_i d_i(d_i+1).
\end{equation}
Thus, for each recommendation we keep only the payoff directions relevant to unilateral deviations, rather than one coordinate for every leaf.

\begin{lemma}\label{lem:projected-regret}
If two distributions on $\prod_i\mathcal A_i$ have equal expectations of $\Psi$, then every unnormalized deviation gain is equal under the two distributions.
\end{lemma}
\begin{proof}
For a deviation from $\sigma_i$ to $\sigma_i'$, write its leaf-realization difference as $\sum_{\ell=1}^{q_i}c_\ell b_{i,\ell}$. By \eqref{eq:feature-regret} and \eqref{eq:projected-features}, its gain under $\mu$ is $\sum_\ell c_\ell\mathbb E_\mu[\Psi_{i,\sigma_i,\ell}]$. Equality follows for every alternative full plan, including plans outside $\mathcal A_i$. Recommendations outside $\mathcal A_i$ have zero gain in both distributions. No equality of the full leaf-contribution expectations is required.
\end{proof}

\paragraph{Feature-expectation polytope.}
For the fixed supports $\mathcal A_i$, define the \emph{feature-expectation polytope}\footnote{This is distinct from the von Stengel--Forges polytope for EFCE \citep{vonstengel2008extensive}, which uses relevant sequence pairs. The feature-expectation polytope depends on the chosen pure-strategy supports and the payoffs. It does not itself impose the NFCE obedience constraints or specify a correlation plan; its purpose is to compress distributions while preserving their deviation gains.}
\begin{equation}\label{eq:feature-polytope}
\mathcal P_\Psi:=\operatorname{conv}\{\Psi(\tau):\tau\in\textstyle\prod_i\mathcal A_i\}
=\{\mathbb E_\mu[\Psi]:\mu\in\Delta(\textstyle\prod_i\mathcal A_i)\}.
\end{equation}
Its affine dimension satisfies
\[
\dim(\mathcal P_\Psi)\le D=\sum_{i=1}^n m_iq_i\le\sum_{i=1}^n d_i(d_i+1),
\]
where the last inequality holds for the sparse mixed strategies in Theorem~\ref{thm:main-two}. We use this polytope to obtain a small-support representation of the product distribution's feature expectation.

\begin{theorem}
\label{thm:leaf-caratheodory}
Every $\phi\in\mathcal P_\Psi$ admits a distribution $\mu_0$ on $\prod_i\mathcal A_i$ such that
\[
\mathbb E_{\mu_0}[\Psi]=\phi,\qquad
|\operatorname{supp}(\mu_0)|\le\dim(\mathcal P_\Psi)+1\le D+1.
\]
In particular, taking $\phi=\mathbb E_{\mu_{\mathrm{prod}}}[\Psi]$ for $\mu_{\mathrm{prod}}=\bigotimes_i\alpha_i$ gives
\[
R_{i,\sigma_i,\sigma_i'}(\mu_0)=R_{i,\sigma_i,\sigma_i'}(\mu_{\mathrm{prod}})
\qquad
\text{for all }i\in[n],\ \sigma_i\in\Sigma_i,\ \sigma_i'\in\Sigma_i.
\]
\end{theorem}

\begin{proof}
By \eqref{eq:feature-polytope}, $\phi$ lies in the convex hull of the feature vectors $\Psi(\tau)$. Applying Carath\'eodory's theorem in the affine hull of $\mathcal P_\Psi$ \citep{rockafellar1970convex}, write $\phi$ as a convex combination of at most $\dim(\mathcal P_\Psi)+1$ such vectors. The corresponding weights on their pure-strategy profiles define $\mu_0$. Since $\mathcal P_\Psi\subseteq\mathbb R^D$, its affine dimension is at most $D$. For the product distribution, its expectation belongs to $\mathcal P_\Psi$ by definition, and the regret equalities follow from Lemma~\ref{lem:projected-regret}, including the zero gains for recommendations outside $\mathcal A_i$.
\end{proof}

For $k\in\{0,1,\ldots,n\}$ and a prefix
$p=(\tau_1,\ldots,\tau_k)\in\prod_{j=1}^k\mathcal A_j$, define
\begin{equation}
\label{eq:phi-prefix}
\phi(p):=\mathbb E\bigl[\Psi(\tau_1,\ldots,\tau_k,\sigma_{k+1},\ldots,\sigma_n)\bigr],
\qquad \sigma_j\sim\alpha_j\text{ independently }(j>k),
\end{equation}
and put $\phi^*:=\phi(\varnothing)$. The support construction preserves
this vector throughout the iterations. Its iterative expansion and support reduction parallels the information-set expansion in \citet[Theorem~4.2]{kollermegiddo1996small}, here applied to player prefixes and conditional payoff features.

\begin{algorithm}[htbp]
\caption{Small-support construction from a product distribution}
\label{alg:leaf-compress}
\begin{algorithmic}[1]
\Require Rational mixed strategies $\alpha_i$ with supports $\mathcal A_i\subseteq\Sigma_i$
\State Compute the bases $b_{i,\ell}$ and $D=\sum_i m_iq_i$.
\State Compute $\phi^*=\phi(\varnothing)$.
\State $P\gets\{(\varnothing,1)\}$.
\For{$k=0,1,\ldots,n-1$}
  \State $E\gets\varnothing$.
  \ForAll{$(p,w)\in P$}
    \ForAll{$\sigma_{k+1}\in\mathcal A_{k+1}$}
      \State Add $(p,\sigma_{k+1})$ to $E$ with temporary weight $w\alpha_{k+1}(\sigma_{k+1})$.
    \EndFor
  \EndFor
  \State Compute $\phi(p')$ for every $p'\in E$.
  \State Compute a basic feasible solution $(x_{p'})_{p'\in E}$ of
  \Statex \hspace{1.5em}$\displaystyle \sum_{p'\in E}x_{p'}\phi(p')=\phi^*,\qquad
  \sum_{p'\in E}x_{p'}=1,\qquad x_{p'}\ge 0.$
  \State $P\gets\{(p',x_{p'}):p'\in E,\ x_{p'}>0\}$.
\EndFor
\State \Return $\mu_0:=\sum_{(\tau,w)\in P}w\,\delta_\tau$.
\end{algorithmic}
\end{algorithm}

\begin{lemma}\label{lem:prefix-eval}
For a prefix $p=(\tau_1,\ldots,\tau_k)$, write $a_{i,\sigma_i,z}(p)$ for the expected leaf contribution $\Phi_{i,\sigma_i,z}$ under the fixed prefix and independent remaining plans. Then
\begin{align}
\label{eq:phi-prefix-eval}
a_{i,\sigma_i,z}(p)
&=p_c(z)u_i(z)
\begin{cases}
\mathbf1\{\tau_i=\sigma_i\},&i\le k,\\
\alpha_i(\sigma_i),&i>k
\end{cases}\notag\\[-1mm]
&\quad\cdot\prod_{\substack{j\ne i\\j\le k}}\chi_j(\tau_j,z)
\prod_{\substack{j\ne i\\j>k}}\left(\sum_{\sigma_j\in\mathcal A_j}\alpha_j(\sigma_j)\chi_j(\sigma_j,z)\right).
\end{align}
The projected prefix expectation has coordinates
\[
\phi_{i,\sigma_i,\ell}(p)=\sum_{z\in Z} b_{i,\ell}(z)a_{i,\sigma_i,z}(p).
\]
In particular, $\phi(p)$ is computable in polynomial bit time.
\end{lemma}
\begin{proof}
In the expectation defining $a_{i,\sigma_i,z}(p)$, the first $k$ plans are fixed and the remaining plans are sampled independently. Substituting \eqref{eq:leaf-Phi}, the recommendation indicator has expectation $\mathbf1\{\tau_i=\sigma_i\}$ if $i\le k$, and $\alpha_i(\sigma_i)$ otherwise. Each opponent's leaf indicator is constant if that opponent is fixed, and otherwise has expectation $\sum_{\sigma_j\in\mathcal A_j}\alpha_j(\sigma_j)\chi_j(\sigma_j,z)$. By independence the expectation of the product equals the product of these expectations, giving \eqref{eq:phi-prefix-eval}. The projected expectation follows by linearity. Each sum and product has polynomially many rational terms of polynomial bit length, including the basis coefficients.
\end{proof}

\begin{lemma}
\label{lem:compress-correct}
Algorithm~\ref{alg:leaf-compress} outputs a rational distribution
$\mu_0$ satisfying
\[
\mathbb E_{\mu_0}[\Psi]=\phi^*
\qquad\text{and}\qquad
|\operatorname{supp}(\mu_0)|\le \dim(\mathcal P_\Psi)+1\le D+1.
\]
The algorithm runs in polynomial bit time.
\end{lemma}

\begin{proof}
At the beginning of iteration $k+1$, maintain the invariant
\[
|P|\le \dim(\mathcal P_\Psi)+1\le D+1,
\qquad
\sum_{(p,w)\in P}w=1,
\qquad
\sum_{(p,w)\in P}w\,\phi(p)=\phi^*.
\]
Initially $P=\{(\varnothing,1)\}$, so the invariant holds because
$\phi(\varnothing)=\phi^*$.

Assume the invariant holds at the beginning of iteration $k+1$.
For every retained prefix $p$,
\[
\phi(p)=\sum_{\sigma_{k+1}\in\mathcal A_{k+1}}\alpha_{k+1}(\sigma_{k+1})\phi(p,\sigma_{k+1})
\]
by conditioning on player $k+1$'s sampled full plan. Therefore the
temporary weights on the expanded family $E$ are nonnegative, sum to
one, and satisfy
\[
\sum_{p'\in E}\widetilde w_{p'}\phi(p')=\phi^*.
\]
Hence the linear system
\[
\sum_{p'\in E}x_{p'}\phi(p')=\phi^*,
\qquad
\sum_{p'\in E}x_{p'}=1,
\qquad x_{p'}\ge 0
\]
is feasible. Every $\phi(p')$ belongs to $\mathcal P_\Psi$ by \eqref{eq:feature-polytope}, since it is the expectation of $\Psi$ under a distribution on $\prod_i\mathcal A_i$. Thus the columns $(\phi(p'),1)$ span a linear space of dimension at most $\dim(\mathcal P_\Psi)+1$: subtracting any one such column from all the others gives vectors in the direction space of the affine hull, with last coordinate zero. The positive columns of a basic feasible solution are linearly independent, so there are at most $\dim(\mathcal P_\Psi)+1$ positive variables. Replacing $P$ by the
positive-support pairs of such a solution preserves the invariant.

After the last iteration, every retained prefix is a full profile
$\tau\in\prod_i\mathcal A_i$, and then $\phi(\tau)=\Psi(\tau)$. Therefore
\[
\mathbb E_{\mu_0}[\Psi]=\sum_{(\tau,w)\in P}w\Psi(\tau)=\phi^*.
\]
The support bound is exactly the basic-feasible-solution bound.

For complexity, at each iteration the family $E$ has size at most
$(D+1)m_{k+1}$. By Lemma~\ref{lem:prefix-eval}, every column $\phi(p')$
is computable in polynomial bit time. The entries of each column and of $\phi^*$ have polynomial bit length: their denominators divide a product of the denominators of the game labels, the original mixed-strategy weights, and the rational basis entries, counted by occurrence. In particular, these coefficients do not depend on weights from preceding support-construction LPs. A basic feasible solution can be computed by exact rational LP in polynomial bit time \citep{gls1988geometric}; its bit length is polynomial by the determinant bound on a nonsingular subsystem. Temporary weights are products of two polynomial-bit rationals. Since there are $n$ iterations, the total running time is polynomial.
\end{proof}

\begin{proof}[Proof of Theorem~\ref{thm:main-two}]
By Lemma~\ref{lem:product-nash-nfce}, the product distribution
$\mu_{\mathrm{prod}}:=\bigotimes_i\alpha_i$ is an unnormalized
$\epsilon$-NFCE. Apply Algorithm~\ref{alg:leaf-compress} to the mixed strategies $\alpha_i$. By Lemma~\ref{lem:compress-correct}, the
output $\mu_0$ satisfies
\[
\mathbb E_{\mu_0}[\Psi]=\mathbb E_{\mu_{\mathrm{prod}}}[\Psi]
\quad\text{and}\quad
|\operatorname{supp}(\mu_0)|\le D+1.
\]
Hence Lemma~\ref{lem:projected-regret} gives
\[
R_{i,\sigma_i,\sigma_i'}(\mu_0)=R_{i,\sigma_i,\sigma_i'}(\mu_{\mathrm{prod}})\le\epsilon
\]
for every $i\in[n]$, every $\sigma_i\in\mathcal A_i$, and every
$\sigma_i'\in\Sigma_i$. If $\sigma_i\notin\mathcal A_i$, then both sides are zero.
Thus $\mu_0$ is an unnormalized $\epsilon$-NFCE. Finally,
\[
|\operatorname{supp}(\mu_0)|\le D+1
=1+\sum_i m_iq_i\le 1+\sum_i d_i(d_i+1)=K.
\]
The polynomial-time bound is provided by
Lemma~\ref{lem:compress-correct}.
\end{proof}

\subsection{Final LP for NFCE}
\label{sec:final-lp}

We use the standard perfect-recall best-response decomposition underlying the sequence form \citep{kollermegiddo1992complexity,kollervonstengel1996efficient}, expressed directly in terms of $\Phi$ and $\chi$.

Let $S\subseteq\Sigma$ be a nonempty finite support. For each player, take $\mathcal A_i=\{\tau_i:\tau\in S\}$ when defining $\Phi$. Let $\mu$ have variables $\mu(\tau)$ for $\tau\in S$, with zero weight outside $S$. The conceptual reweighting LP assumes terminal payoffs in $[0,1]$, so that $t=1$ is feasible for any probability weights (the exact feasibility system at $t=0$ needs no such normalization):
\begin{equation}\label{eq:conceptual-lp}
\begin{aligned}
 t^*(S)=\min\quad&t\\
 \text{subject to}\quad
 &\sum_{\tau\in S}\mu(\tau)=1,\qquad \mu(\tau)\ge0,\qquad 0\le t\le1,\\
 &R_{i,\sigma_i,\sigma_i'}(\mu)\le t
 \qquad(i\in[n],\ \sigma_i\in\mathcal A_i,\ \sigma_i'\in\Sigma_i).
\end{aligned}
\end{equation}
The regret constraints are linear by \eqref{eq:feature-regret}. Only the alternative strategies may require exponentially many constraints.

For every information set $I\in\mathcal I_i$, let $\operatorname{par}_i(I)$ be its last preceding own information-set/action pair, or $\bot$ if none. For a leaf $z$, let $\operatorname{last}_i(z)$ be its last own information-set/action pair, or $\bot$ if none.
For each $i,\sigma_i\in\mathcal A_i,I\in\mathcal I_i$, introduce an unrestricted variable $W_{i,\sigma_i,I}$ and impose
\begin{align}
\label{eq:dp-constraints}
W_{i,\sigma_i,I}&\ge
\sum_{z:\operatorname{last}_i(z)=(I,e)}\mathbb E_\mu[\Phi_{i,\sigma_i,z}]
+\sum_{J:\operatorname{par}_i(J)=(I,e)}W_{i,\sigma_i,J}
\quad(e\in A(I)),\\
\label{eq:dp-root}
\sum_{z:\operatorname{last}_i(z)=\bot}\mathbb E_\mu[\Phi_{i,\sigma_i,z}]
&+\sum_{I:\operatorname{par}_i(I)=\bot}W_{i,\sigma_i,I}
\le\sum_{z\in Z}\chi_i(\sigma_i,z)\mathbb E_\mu[\Phi_{i,\sigma_i,z}]+t.
\end{align}
Here every $\mathbb E_\mu[\Phi_{i,\sigma_i,z}]$ is the linear expression $\sum_{\tau\in S}\mu(\tau)\Phi_{i,\sigma_i,z}(\tau)$, not an additional variable. The compact LP minimizes $t$ with the normalization and nonnegativity constraints of \eqref{eq:conceptual-lp}, the bounds $0\le t\le1$, and \eqref{eq:dp-constraints}--\eqref{eq:dp-root}.

\begin{lemma}\label{lem:compact-lp}
The projection of the compact LP onto the distribution weights and $t$ is exactly \eqref{eq:conceptual-lp}. Setting $t=0$ gives a polynomial-size rational LP for the exact NFCE distributions supported on $S$.
\end{lemma}
\begin{proof}
Fix $i$, $\sigma_i\in\mathcal A_i$ and the probability weights $\mu$. By \eqref{eq:feature-regret}, the maximum regret over all alternative full plans equals
\[
\max_{\sigma_i'\in\Sigma_i}\sum_{z\in Z}\chi_i(\sigma_i',z)\mathbb E_\mu[\Phi_{i,\sigma_i,z}]
-\sum_{z\in Z}\chi_i(\sigma_i,z)\mathbb E_\mu[\Phi_{i,\sigma_i,z}].
\]
For each $I\in\mathcal I_i$, define the optimal continuation value $V_I$ recursively, in reverse order of own-history length, by
\[
V_I=\max_{e\in A(I)}\left\{\sum_{z:\operatorname{last}_i(z)=(I,e)}\mathbb E_\mu[\Phi_{i,\sigma_i,z}]
+\sum_{J:\operatorname{par}_i(J)=(I,e)}V_J\right\}.
\]
Conditional on compatibility with the preceding own history, a full plan selects one action $e$ at $I$. It includes the contributions of leaves ending in $(I,e)$, and the contributions from the child information sets $J$ with $\operatorname{par}_i(J)=(I,e)$. Perfect recall implies that distinct child families have disjoint descendant information sets, since each information set has a unique preceding own history. Choices in these families can therefore be optimized independently. Induction on the remaining own-history length proves the recursion. Leaves with no own action contribute independently of the chosen plan; the maximum in the preceding display for regret is consequently attained with alternative payoff
\[
\sum_{z:\operatorname{last}_i(z)=\bot}\mathbb E_\mu[\Phi_{i,\sigma_i,z}]
+\sum_{I:\operatorname{par}_i(I)=\bot}V_I.
\]
Reverse induction in \eqref{eq:dp-constraints} gives $W_{i,\sigma_i,I}\ge V_I$ for every $I$. Thus \eqref{eq:dp-root} implies that the maximum regret is at most $t$. Conversely, if the maximum regret is at most $t$, choosing $W_{i,\sigma_i,I}=V_I$ satisfies both sets of constraints. This applies to every supported recommendation; all other recommendations have zero regret. It proves the claimed equivalence. The number of variables and constraints and their rational encoding lengths are polynomial in the explicit game size and $|S|$, since each expectation is a sum over $S$ and each coefficient is computed by \eqref{eq:leaf-Phi}.
\end{proof}

\subsection{Choosing the \texorpdfstring{$\epsilon$}{epsilon} carefully}
\label{sec:choosing-epsilon}

We now prove Theorem~\ref{thm:exact-support}.

Let $u_{\min}$ and $u_{\max}$ be the minimum and maximum terminal
payoffs among all players. If $u_{\max}>u_{\min}$, replace every
terminal payoff by
\[
\widetilde u_i(z):=\frac{u_i(z)-u_{\min}}{u_{\max}-u_{\min}}\in[0,1].
\]
If $u_{\max}=u_{\min}$, replace every payoff by $0$. In the rest of this section and Appendix~\ref{app:gap}, we assume this normalization has been performed and write $u_i(z)$ for the normalized payoffs.

Let $\mathcal R$ contain all normalized terminal payoffs and all
chance-edge probabilities, counting each occurrence in the input, and
set
\[
B:=\prod_{r\in\mathcal R}\operatorname{den}(r).
\]
Put
\[
N:=|Z|,
\qquad
K:=1+\sum_i d_i(d_i+1),
\qquad
\eta:=((K+1)B)^{-(K+1)}.
\]

\begin{lemma}
\label{lem:uniform-gap}
For every nonempty support $S\subseteq\Sigma$ with $|S|\le K$,
\[
t^*(S)=0
\qquad\text{or}\qquad
t^*(S)\ge \eta.
\]
Moreover, $\log(1/\eta)$ is polynomial in the encoding length of the
input game.
\end{lemma}

The proof is given in Appendix~\ref{app:gap}.

\begin{proof}[Proof of Theorem~\ref{thm:exact-support}]
Set
\[
\epsilon_\Gamma:=\frac{\eta}{2}=\frac{1}{2((K+1)B)^{K+1}}.
\]
Let $\mu$ be an unnormalized $\epsilon_\Gamma$-NFCE supported on
$S=\{\tau^1,\ldots,\tau^k\}$ with $k\le K$. Use its probabilities as the weight variables in \eqref{eq:conceptual-lp}, and set
$t:=\epsilon_\Gamma$. By definition of unnormalized
$\epsilon_\Gamma$-NFCE, all deviation inequalities of
\eqref{eq:conceptual-lp} hold. Therefore
\[
t^*(S)\le \epsilon_\Gamma=\frac{\eta}{2}<\eta.
\]
By Lemma~\ref{lem:uniform-gap}, this is possible only if
$t^*(S)=0$. Hence the compact LP of Section~\ref{sec:final-lp} with
$t=0$ is feasible by Lemma~\ref{lem:compact-lp}. Any feasible
solution yields an exact NFCE supported on $S$; removing zero-weight
profiles yields an exact NFCE $\mu^*$ with
$\operatorname{supp}(\mu^*)\subseteq S$.
\end{proof}

\begin{proof}[Proof of Theorem~\ref{thm:main-upper}]
Normalize the game as above. If the payoffs are not all equal, a common positive affine transformation multiplies every deviation gain by the same positive scalar; if they are all equal, every profile is already an NFCE. Thus exact NFCEs of the normalized game are exact NFCEs of the original game. Zero-probability chance edges may be pruned if required by the approximate-Nash oracle; pure plans can then be extended arbitrarily on deleted information sets, since those branches are unreachable under every deviation. If necessary, inactive players can be added for the oracle call and removed afterwards. If the oracle requires positive integer terminal payoffs, clear denominators and add a common constant, scaling the requested regret tolerance by the same positive multiplicative factor. These operations have polynomial bit complexity.

Compute $N$, $K$, $B$, and $\epsilon_\Gamma$. By Theorem~\ref{thm:approx-nash-ppad}, obtain a rational behavioral weak $\epsilon_\Gamma$-Nash equilibrium using one PPAD search reduction. Apply the small-support realization-equivalent decomposition described in Section~\ref{sec:overview-ppad} to obtain rational mixed strategies with $m_i\le d_i+1$. By realization equivalence, the resulting mixed profile is still weak $\epsilon_\Gamma$-Nash. Lemma~\ref{lem:product-nash-nfce} gives an unnormalized $\epsilon_\Gamma$-NFCE on their independent product. Theorem~\ref{thm:main-two} computes in polynomial bit time a distribution $\mu_0$ preserving all deviation gains, with support size at most $K$. By Theorem~\ref{thm:exact-support} and Lemma~\ref{lem:compact-lp}, the final compact LP with $t=0$ is feasible and an exact rational solution can be computed in polynomial bit time \citep{gls1988geometric}. Its profile weights have polynomial encoding length. Since $\sum_i d_i\le N-1$ by Section~\ref{sec:overview-ppad}, $K=1+\sum_i(d_i^2+d_i)\le 1+(N-1)^2+(N-1)=1+N(N-1)$.

All postprocessing is polynomial bit time and succeeds for every valid oracle answer. Given a candidate rational output of polynomial length, checking normalization and computing the maximum deviation for each supported recommendation using the dynamic program in Lemma~\ref{lem:compact-lp} verifies exact NFCE in polynomial time. Thus, with a polynomial output-encoding bound sufficient for the rational LP solution, the search relation is polynomially balanced and verifiable. Composing the reduction with the weak approximate-Nash reduction proves PPAD membership.
\end{proof}

\begin{proof}[Proof of Corollary~\ref{cor:existence}]
The search reduction in Theorem~\ref{thm:main-upper} is total and produces a rational output of the stated support and encoding bounds.
\end{proof}

\section{PPAD-hardness over quadratic support}
\label{sec:hardness}
\subsection{Overview}
\label{sec:hard-overview}
In this section, we provide an overview of the hardness reduction for exact NFCE computation. Our reduction is from the problem of computing a weak approximate Nash equilibrium.

\begin{theorem}[Approximate bimatrix Nash is PPAD-hard~[\cite{chen2009settling}]
\label{thm:source-nash-hardness}
Given a rational payoff matrix $P,Q\in[0,1]^{r\times r}$, computing a weak $1/r$-Nash equilibrium of the corresponding bimatrix game is $\mathrm{PPAD}$-hard.
\end{theorem}

We combine the source game (branch $G$) with a constant-sum matching game (branch $M$), selected by a public fair coin. The latter is the standard match/mismatch game used, for example, in \citet[Example~3]{babichenko2014small}, implemented here by a binary tree. Let $m=2^k$, where $k\ge1$ is chosen in Section~\ref{sec:hard-parameters}. For a full profile $\tau$, write $X_i=X_i(\tau_i)\in[m]$ for player $i$'s induced matching label and $Y_i=Y_i(\tau_i)\in[r]$ for its source-game action. These maps are specified in Section~\ref{sec:hard-reduction}. Under a distribution $\mu$ on full profiles, all these variables are sampled using $\tau\sim\mu$. Let $\pi$ be the law of $(Y_1,Y_2)$, with marginals $x,y$. We write $\mathsf H$ for Shannon entropy and $\mathsf I$ for mutual information.

\begin{lemma}\label{lem:hard-construction}
The constructed game has perfect recall and satisfies
\begin{align*}
|V|&=6m+r^2+r-3, & N&=3m-2+r^2,\\
d_1=d_2&=m+r-2, & |\Sigma_1|=|\Sigma_2|&=r\,2^{m-1}.
\end{align*}
For every full profile $\tau$, with $X_i=X_i(\tau_i)$ and $Y_i=Y_i(\tau_i)$, the chance-averaged payoffs are
\begin{align}
U_1(\tau)&=\tfrac12 P_{Y_1Y_2}+\tfrac12\mathbf1\{X_1=X_2\},\label{eq:hard-u1}\\
U_2(\tau)&=\tfrac12 Q_{Y_1Y_2}+\tfrac12\mathbf1\{X_1\ne X_2\}.\label{eq:hard-u2}
\end{align}
Each player can replace its matching label by any element of $[m]$ while keeping its branch-$G$ action unchanged, and can replace its branch-$G$ action while leaving its entire matching plan unchanged. The game admits an exact rational NFCE with support at most $\kappa_d=(m+r-1)^2$.
The proof is given in Section~\ref{sec:hard-reduction}.
\end{lemma}

The matching incentives force uncertainty about the opponent's matching label even conditional on one's entire recommendation. The next two lemmas use standard entropy identities and Pinsker's inequality \citep{cover2006elements,pinsker1964information}, collected in Appendix~\ref{app:information}.

\begin{lemma}\label{lem:matching-entropy}
If $\mu$ is an exact NFCE, then conditional on either player's full recommendation of positive probability, the opponent's matching label is uniform on $[m]$. In particular,
\begin{equation}\label{eq:matching-entropy}
\mathsf H(X_2\mid\tau_1)=\mathsf H(X_1\mid\tau_2)=\log m.
\end{equation}
The proof is given in Section~\ref{sec:hard-entropy}.
\end{lemma}

The support constraint bounds the joint entropy, leaving little room for dependence between the source-game actions.

\begin{lemma}\label{lem:near-independent}
Let $\kappa$ be any positive integer support size bound. If $\mu$ is an exact NFCE with $|\operatorname{supp}(\mu)|\le\kappa$, then $\kappa\ge m^2$ and
\begin{equation}\label{eq:mi-bound}
\mathsf I(Y_1;Y_2)\le\log(\kappa/m^2),
\end{equation}
and
\begin{equation}\label{eq:tv-bound}
\operatorname{TV}(\pi,x\otimes y)\le
\sqrt{\frac{\log(\kappa/m^2)}{2}}.
\end{equation}
The proof is given in Section~\ref{sec:hard-entropy}.
\end{lemma}

Near-independence allows the source-game marginals to serve as the Nash decoder.

\begin{lemma}\label{lem:marginal-decoder}
Under the hypotheses of Lemma~\ref{lem:near-independent}, the marginal strategy profile $(x,y)$ of the source game is a weak $\zeta$-Nash equilibrium, where
\[
\zeta=\sqrt{\frac{\log(\kappa/m^2)}{2}}.
\]
The proof is given in Section~\ref{sec:hard-decoder}.
\end{lemma}

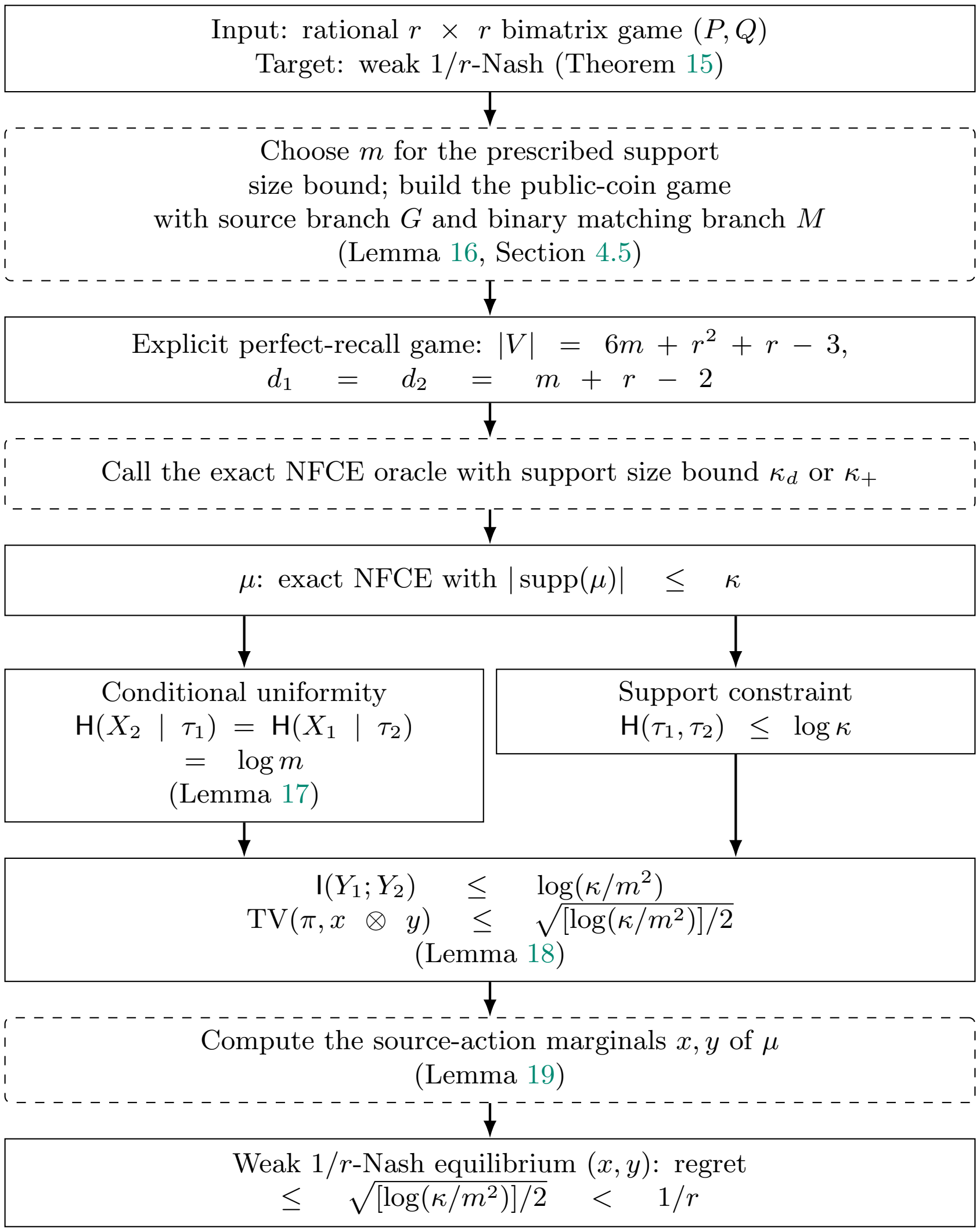
\begin{figure}[htbp]
\centering
\begin{tikzpicture}[
 font=\footnotesize,
 procedure/.style={draw,dashed,rounded corners=2pt,align=center,
   inner xsep=5pt,inner ysep=4pt,minimum height=8mm},
 object/.style={draw,solid,align=center,
   inner xsep=5pt,inner ysep=4pt,minimum height=8mm},
 flow/.style={-{Latex[length=2mm]},thick},
 node distance=4mm
]
\node[object,text width=10.7cm] (source)
 {Input: rational $r\times r$ bimatrix game $(P,Q)$\\
  Target: weak $1/r$-Nash (Theorem~\ref{thm:source-nash-hardness})};
\node[procedure,text width=10.7cm,below=of source] (construct)
 {Choose $m$ for the prescribed support size bound; build the public-coin game\\
  with source branch $G$ and binary matching branch $M$\\
  (Lemma~\ref{lem:hard-construction}, Section~\ref{sec:hard-parameters})};
\node[object,text width=10.7cm,below=of construct] (game)
 {Explicit perfect-recall game: $|V|=6m+r^2+r-3$,\\
  $d_1=d_2=m+r-2$};
\node[procedure,text width=10.7cm,below=of game] (oracle)
 {Call the exact NFCE oracle with support size bound $\kappa_d$ or $\kappa_+$};
\node[object,text width=10.7cm,below=of oracle] (mu)
 {$\mu$: exact NFCE with $|\operatorname{supp}(\mu)|\le\kappa$};
\node[object,text width=5.1cm,below=6mm of mu.south west,anchor=north west] (entropy)
 {Conditional uniformity\\
  $\mathsf H(X_2\mid\tau_1)=\mathsf H(X_1\mid\tau_2)$\\
  $=\log m$\\
  (Lemma~\ref{lem:matching-entropy})};
\node[object,text width=5.1cm,below=6mm of mu.south east,anchor=north east] (support)
 {Support constraint\\
  $\mathsf H(\tau_1,\tau_2)\le\log\kappa$};
\node[object,text width=10.7cm,anchor=north] (info) at ([yshift=-4mm]entropy.south -| mu.center)
 {$\mathsf I(Y_1;Y_2)\le\log(\kappa/m^2)$\\
  $\operatorname{TV}(\pi,x\otimes y)\le\sqrt{[\log(\kappa/m^2)]/2}$\\
  (Lemma~\ref{lem:near-independent})};
\node[procedure,text width=10.7cm,below=of info] (decode)
 {Compute the source-action marginals $x,y$ of $\mu$\\
  (Lemma~\ref{lem:marginal-decoder})};
\node[object,text width=10.7cm,below=of decode] (output)
 {Weak $1/r$-Nash equilibrium $(x,y)$: regret\\
  $\le \sqrt{[\log(\kappa/m^2)]/2}<1/r$};
\draw[flow] (source)--(construct);
\draw[flow] (construct)--(game);
\draw[flow] (game)--(oracle);
\draw[flow] (oracle)--(mu);
\draw[flow] (mu.south -| entropy.north)--(entropy.north);
\draw[flow] (mu.south -| support.north)--(support.north);
\draw[flow] (entropy.south)--(info.north -| entropy.south);
\draw[flow] (support.south) -- (info.north -| support.south);
\draw[flow] (info)--(decode);
\draw[flow] (decode)--(output);
\end{tikzpicture}
\caption{Hardness proof roadmap. The entropy and support bounds certify the decoder; the decoder itself only computes marginals.}
\label{fig:hardness-overview}
\end{figure}

For the dimension-based support size bound we choose $m=2^{\lceil\log_2(8r^3)\rceil}$; for the node-based support size bound we choose $m=2^{\lceil\log_2(8r^4)\rceil}$ (with $r\ge5$). Section~\ref{sec:hard-parameters} verifies that the resulting regret is below $1/r$, that the construction is polynomial in size, and that the stated support size bounds are nonvacuous. It also establishes PPAD membership for the exact search problems. Figure~\ref{fig:hardness-overview} summarizes the reduction and its decoding argument.

\subsection{The reduction}
\label{sec:hard-reduction}
Let $P,Q\in[0,1]^{r\times r}$ be the rational payoff matrices of the source game, with $r\ge2$. Let $m=2^k$ for an integer $k\ge1$, to be chosen below. Construct a two-player extensive-form game $\Gamma$ with a public chance move selecting branch $G$ or branch $M$, each with probability $1/2$.

In branch $G$, player~1 selects an action in $[r]$, then player~2 selects an action in $[r]$ without observing player~1's action; the payoffs are given by $P,Q$. The $r$ nodes of player~2 in this branch form a single information set.

\begin{figure}[htbp]
\centering
\begin{minipage}{0.94\linewidth}
\centering
\small
\textbf{Example source game ($r=2$)}\quad
$\begin{array}{c|cc}
 &1&2\\ \hline
1&(1,1)&(0,\tfrac12)\\
2&(\tfrac12,0)&(\tfrac34,\tfrac34)
\end{array}$
\end{minipage}
\par\medskip
\begin{tikzpicture}[x=0.94cm,y=0.88cm,font=\scriptsize,
  player/.style={circle,draw,fill=white,inner sep=1pt,minimum size=4.8mm},
  chance/.style={regular polygon,regular polygon sides=3,draw=black,fill=white,inner sep=1pt,minimum size=5.8mm,text=black},
  iset/.style={player,draw=#1!85!black,fill=#1!10,text=#1!85!black},
  ilink/.style={densely dashed,#1!85!black,thick},
  leaf/.style={inner sep=1pt,fill=white},
  ed/.style={thin}]
\node[chance] (c) at (5.5,0) {$c$};
\node[player] (g) at (1.8,-1.35) {$1$};
\node[player] (m) at (9,-1.35) {$1$};
\draw[ed] (c)--node[above left] {$1/2$} (g);
\draw[ed] (c)--node[above right] {$1/2$} (m);
\node[above=2pt of g] {\textbf{$G$}};
\node[above=2pt of m] {\textbf{$M$} ($k=2$)};
\node[iset=BrickRed] (g1) at (.8,-2.65) {$2$};
\node[iset=BrickRed] (g2) at (2.8,-2.65) {$2$};
\draw[ed] (g)--node[above left] {$1$} (g1);
\draw[ed] (g)--node[above right] {$2$} (g2);
\draw[ilink=BrickRed] (g1)--(g2);
\node[leaf] (ga) at (0,-4.0) {$(1,1)$};
\node[leaf] (gb) at (1.25,-4.0) {$(0,\frac12)$};
\node[leaf] (gc) at (2.35,-4.0) {$(\frac12,0)$};
\node[leaf] (gd) at (3.65,-4.0) {$(\frac34,\frac34)$};
\draw[ed] (g1)--node[left] {$1$} (ga);
\draw[ed] (g1)--node[right] {$2$} (gb);
\draw[ed] (g2)--node[left] {$1$} (gc);
\draw[ed] (g2)--node[right] {$2$} (gd);
\node[iset=RoyalBlue] (m0) at (7,-2.65) {$2$};
\node[iset=RoyalBlue] (m1) at (11,-2.65) {$2$};
\draw[ed] (m)--node[above left] {$0$} (m0);
\draw[ed] (m)--node[above right] {$1$} (m1);
\draw[ilink=RoyalBlue] (m0)--(m1);
\node[player] (p0) at (6.1,-4.0) {$1$};
\node[leaf] (f01) at (7.85,-4.0) {$(0,1)$};
\node[leaf] (f10) at (10.15,-4.0) {$(0,1)$};
\node[player] (p1) at (11.9,-4.0) {$1$};
\draw[ed] (m0)--node[left] {$0$} (p0);
\draw[ed] (m0)--node[right] {$1$} (f01);
\draw[ed] (m1)--node[left] {$0$} (f10);
\draw[ed] (m1)--node[right] {$1$} (p1);
\foreach \prefix/\xx/\isetcolor in {a/6.1/ForestGreen,b/11.9/Purple}{
  \node[iset=\isetcolor] (\prefix L) at (\xx-1.02,-5.35) {$2$};
  \node[iset=\isetcolor] (\prefix R) at (\xx+1.02,-5.35) {$2$};
  \draw[ilink=\isetcolor] (\prefix L)--(\prefix R);
  \node[leaf] (\prefix 00) at (\xx-1.68,-6.7) {$(1,0)$};
  \node[leaf] (\prefix 01) at (\xx-.56,-6.7) {$(0,1)$};
  \node[leaf] (\prefix 10) at (\xx+.56,-6.7) {$(0,1)$};
  \node[leaf] (\prefix 11) at (\xx+1.68,-6.7) {$(1,0)$};
  \draw[ed] (\prefix L)--node[left] {$0$} (\prefix 00);
  \draw[ed] (\prefix L)--node[right] {$1$} (\prefix 01);
  \draw[ed] (\prefix R)--node[left] {$0$} (\prefix 10);
  \draw[ed] (\prefix R)--node[right] {$1$} (\prefix 11);
}
\draw[ed] (p0)--node[above left] {$0$} (aL);
\draw[ed] (p0)--node[above right] {$1$} (aR);
\draw[ed] (p1)--node[above left] {$0$} (bL);
\draw[ed] (p1)--node[above right] {$1$} (bR);
\end{tikzpicture}
\caption{Illustrative reduction of a $2\times2$ bimatrix game with four matching labels ($m=4$, $k=2$). The left branch implements the payoff matrix above; the right branch is the entire binary matching tree. Each colored dashed line joins nodes in one non-singleton information set; different colors distinguish the four such sets. The triangle denotes chance. The choice $m=4$ is solely for illustrating the construction, not the hardness parameters. The tree has $27$ nodes.}
\label{fig:hardness-game}
\end{figure}
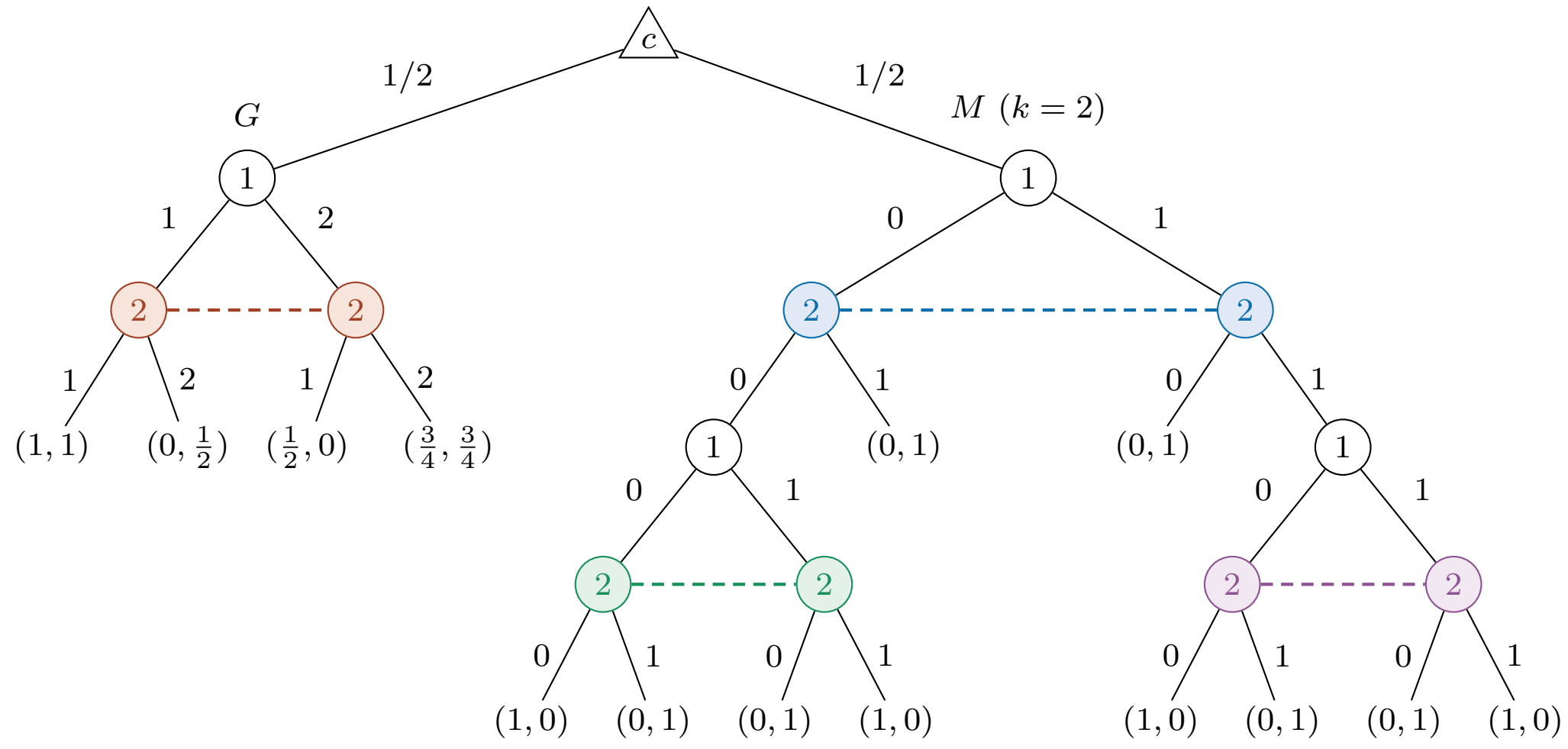

In branch $M$, identify $[m]$ with the binary strings of length $k$. At each common prefix $w\in\{0,1\}^{<k}$, player~1 chooses a bit, then player~2 chooses a bit without observing player~1's current bit. The two nodes of player~2 at prefix $w$ form one information set. Different prefixes correspond to different information sets. If the bits disagree, play ends at payoff $(0,1)$. If they agree on bit $e$, play continues at prefix $we$, or ends at payoff $(1,0)$ if $|we|=k$. Both players know the branch and the previous agreed bits. Figure~\ref{fig:hardness-game} illustrates the construction for a small bimatrix game and four matching labels.

For a full plan $\sigma_i\in\Sigma_i$, let $Y_i(\sigma_i)\in[r]$ be its branch-$G$ action. Define its matching label $X_i(\sigma_i)\in[m]$ by starting at the empty prefix and recursively following that player's prescribed bit at each resulting prefix until $k$ bits have been selected. These are functions of the full plan, not a replacement of its recommendation label. The branch coin is independent of the sampled full profile $\tau\sim\mu$.

\begin{proof}[Proof of Lemma~\ref{lem:hard-construction}]
Branch $G$ has $1+r+r^2$ nodes and $r^2$ leaves. In branch $M$, the number of agreed prefixes of length less than $k$ is $\sum_{j=0}^{k-1}2^j=m-1$. At each such prefix there is one player~1 node, two player~2 nodes and two disagreement leaves. There are also $m$ final agreement leaves. Thus branch $M$ has $5(m-1)+m=6m-5$ nodes and $2(m-1)+m=3m-2$ leaves. Adding the public chance node gives the stated counts. The rational labels have polynomial encoding length.

The information sets of player~1 are singletons. The two nodes of player~2 in a matching information set have the same previous own information sets and bits, since they differ only in player~1's current bit. The same perfect-recall condition holds in branch $G$, where player~2 has not previously moved. Both players have one information set with $r$ actions and $m-1$ binary information sets. Therefore the dimension formula of Section~\ref{sec:overview-ppad} yields $d_i=r-1+(m-1)$, and the number of full plans is $r2^{m-1}$.

For each fixed pair of plans, while the matching labels have a common prefix, play follows that prefix: this follows by induction on its length from the definition of $X_i$. If the labels first differ at some bit, the game terminates there with $(0,1)$; otherwise it reaches a full agreement leaf with $(1,0)$. Averaging over the public chance move proves \eqref{eq:hard-u1}--\eqref{eq:hard-u2}. To implement a prescribed matching label, set the bits at its successive prefixes to its successive bits and choose arbitrary bits elsewhere, without changing the source action. To change only the source action, keep all matching information-set choices fixed. These are valid full-plan replacements. Finally, take a Nash equilibrium of the two-player game and decompose its realization plans into mixed strategies of support at most $d_i+1$ using the dimension-based sparsification in Section~\ref{sec:overview-ppad}. The realization plans can be chosen rational by the two-player sequence-form formulation \citep{kollervonstengel1996efficient}; their independent product is an exact rational NFCE supported on at most $\kappa_d$ profiles.
\end{proof}

\subsection{Matching incentives and the support constraint}
\label{sec:hard-entropy}
For each positive-probability full recommendation, define
\[
M(\sigma_1):=\max_{j\in[m]}\Pr(X_2=j\mid\tau_1=\sigma_1),\qquad
L(\sigma_2):=\min_{j\in[m]}\Pr(X_1=j\mid\tau_2=\sigma_2).
\]
Values on null recommendations are immaterial.

\begin{proof}[Proof of Lemma~\ref{lem:matching-entropy}]
Let $p=\Pr(X_1=X_2)$. For every supported recommendation $\sigma_1\in\Sigma_1$, player~1 may replace its matching plan by one implementing a label attaining $M(\sigma_1)$, keeping its source action unchanged (Lemma~\ref{lem:hard-construction}). This defines a deterministic deviation rule, extended arbitrarily on null recommendations. The branch-$G$ payoff is unchanged; exact obedience in \eqref{eq:global-nfce} and the payoff formula \eqref{eq:hard-u1} imply $p\ge\mathbb E[M(\tau_1)]$, after cancelling the common factor $1/2$.
Similarly, player~2 can implement a label attaining $L(\sigma_2)$ without changing its source action. Exact obedience and \eqref{eq:hard-u2} give $p\le\mathbb E[L(\tau_2)]$. Since the maximum of $m$ probabilities is at least $1/m$ and their minimum is at most $1/m$,
\[
\frac1m\le\mathbb E[M(\tau_1)]\le p\le\mathbb E[L(\tau_2)]\le\frac1m.
\]
All inequalities are equalities. The pointwise nonnegative quantities $M(\tau_1)-1/m$ and $1/m-L(\tau_2)$ therefore have expectation zero, so both vanish on every supported recommendation. A distribution on $[m]$ whose maximum (or minimum) is $1/m$ must be uniform. This proves conditional uniformity and \eqref{eq:matching-entropy}.
\end{proof}

\begin{proof}[Proof of Lemma~\ref{lem:near-independent}]
The joint recommendation has at most $\kappa$ possible values, hence $\mathsf H(\tau_1,\tau_2)\le\log\kappa$. Since $X_i$ is a deterministic function of the full recommendation $\tau_i$, the conditional chain rule implies
\begin{align*}
\mathsf H(\tau_1\mid\tau_2)&\ge\mathsf H(X_1\mid\tau_2),\\
\mathsf H(\tau_2\mid\tau_1)&\ge\mathsf H(X_2\mid\tau_1).
\end{align*}
Thus Lemma~\ref{lem:matching-entropy} yields
\begin{align*}
\mathsf I(\tau_1;\tau_2)
&=\mathsf H(\tau_1,\tau_2)-\mathsf H(\tau_1\mid\tau_2)-\mathsf H(\tau_2\mid\tau_1)\\
&\le\log\kappa-2\log m.
\end{align*}
Since $Y_i$ is a deterministic function of $\tau_i$, the data-processing inequality gives $\mathsf I(Y_1;Y_2)\le\mathsf I(\tau_1;\tau_2)$. This proves \eqref{eq:mi-bound}; nonnegativity of mutual information also gives $\kappa\ge m^2$. Apply Pinsker's inequality (Lemma~\ref{lem:pinsker}) to $\pi$ and $x\otimes y$, using $D_{\rm KL}(\pi\|x\otimes y)=\mathsf I(Y_1;Y_2)$, to obtain \eqref{eq:tv-bound}.
\end{proof}

\subsection{Decoding a Nash equilibrium}
\label{sec:hard-decoder}
\begin{proof}[Proof of Lemma~\ref{lem:marginal-decoder}]
For any $a\in[r]$, player~1 may use the deviation rule that replaces the branch-$G$ action by $a$ in each full recommendation, leaving all matching bits unchanged. By \eqref{eq:global-nfce} and \eqref{eq:hard-u1},
\[
e_a^\top P y-\mathbb E_\pi[P_{Y_1Y_2}]\le0,
\]
where $e_a$ is the $a$th unit vector. Since the entries of $P$ lie in $[0,1]$, Lemma~\ref{lem:pinsker} gives
\[
\mathbb E_\pi[P_{Y_1Y_2}]\le x^\top P y+\operatorname{TV}(\pi,x\otimes y).
\]
Therefore $e_a^\top P y-x^\top P y\le\operatorname{TV}(\pi,x\otimes y)$ for all $a$. Likewise, for every $b\in[r]$, player~2 may replace only its branch-$G$ action by $b$, yielding
\[
x^\top Qe_b-\mathbb E_\pi[Q_{Y_1Y_2}]\le0,
\qquad
\mathbb E_\pi[Q_{Y_1Y_2}]\le x^\top Qy+\operatorname{TV}(\pi,x\otimes y).
\]
Thus $x^\top Qe_b-x^\top Qy\le\operatorname{TV}(\pi,x\otimes y)$. Both players' pure deviations satisfy the bound, and mixed deviations cannot exceed the maximum pure gain. Substitute \eqref{eq:tv-bound} to conclude.
\end{proof}

\subsection{Choice of parameters and complexity}
\label{sec:hard-parameters}
\begin{proof}[Proof of Theorem~\ref{thm:main-hardness}]
By Theorem~\ref{thm:source-nash-hardness}, we may reduce from finding a weak $1/r$-Nash equilibrium in an $r\times r$ rational bimatrix game. For the second reduction we can restrict to $r\ge5$: smaller fixed dimensions can be solved in polynomial bit time by enumerating the finitely many pairs of supports and solving rational linear feasibility problems for the corresponding Nash conditions. Thus this restriction does not affect PPAD-hardness (equivalently, duplicate actions to pad small games).

For either support size bound, Lemma~\ref{lem:marginal-decoder} bounds source-game regret by
\begin{equation}\label{eq:binary-general-regret}
\zeta=\sqrt{\tfrac12\log(\kappa/m^2)}.
\end{equation}

\paragraph{Dimension-based support size bound.}
Set $m=2^{\lceil\log_2(8r^3)\rceil}$, so $8r^3\le m<16r^3$, and $\kappa=\kappa_d=(m+r-1)^2$. Since $\log(1+t)\le t$ for $t\ge0$,
\[
\tfrac12\log(\kappa/m^2)=\log\left(1+\frac{r-1}{m}\right)\le\frac{r-1}{m}.
\]
It follows from \eqref{eq:binary-general-regret} that
\[
\zeta\le\sqrt{\frac{r-1}{m}}\le\frac{1}{\sqrt{8}\,r}<\frac1r,
\]
using $m\ge8r^3$. Lemma~\ref{lem:hard-construction} gives $|V|=\Theta(r^3)$ and existence under $\kappa_d=\Theta(|V|^2)$.

\paragraph{Node-based support size bound.}
Set $m=2^{\lceil\log_2(8r^4)\rceil}$, so $8r^4\le m<16r^4$, and assume $r\ge5$. The node count of Lemma~\ref{lem:hard-construction} gives
\[
|V|-6(m+r-1)=r^2-5r+3\ge0.
\]
Since $(m+r-1)^2$ is an integer, this implies $\kappa_d\le\lfloor |V|^2/36\rfloor=\kappa_V$. Hence $\kappa_+=\kappa_V$ on these instances, and the support size bound admits an exact NFCE. With $\kappa=\kappa_V$, we have
\[
\tfrac12\log(\kappa/m^2)\le\log\left(\frac{|V|}{6m}\right)
=\log\left(1+\frac{r^2+r-3}{6m}\right)\le\frac{r^2+r-3}{6m}.
\]
Substitute into \eqref{eq:binary-general-regret}:
\[
\zeta\le\sqrt{\frac{r^2+r-3}{6m}}\le\frac{1}{\sqrt{24}\,r}<\frac1r,
\]
where $r^2+r-3\le2r^2$ and $m\ge8r^4$. The node count is $\Theta(r^4)$ and $\kappa=\Theta(|V|^2)$.

\paragraph{Polynomial reduction and decoding.}
The game has polynomially many nodes and rational labels of polynomial bit length. A full plan specifies one action at each information set, so its encoding is polynomial in the tree size. Given the explicit distribution with weights $w_\ell$ on full profiles $\sigma^\ell$, the decoder computes
\[
x_a=\sum_{\ell:Y_1(\sigma_1^\ell)=a}w_\ell,\qquad
y_b=\sum_{\ell:Y_2(\sigma_2^\ell)=b}w_\ell.
\]
The sums are computable in polynomial bit time because the supported profiles and rational weights have polynomial encoding length. Every valid oracle answer is an exact NFCE within the specified support size bound and yields a weak $1/r$-Nash equilibrium of the source game. This proves PPAD-hardness of both prescribed-support exact search problems.

\paragraph{PPAD membership.}
On every two-player input, use the weak approximate Nash oracle and the dimension-based decompositions as in the proof of Theorem~\ref{thm:main-upper}, but form the product explicitly instead of compressing it. Its support contains at most $\kappa_d=(d_1+1)(d_2+1)\le K$ profiles, since $K-\kappa_d=d_1^2+d_2^2-d_1d_2\ge0$. Lemma~\ref{lem:product-nash-nfce} gives the required unnormalized approximate NFCE and Theorem~\ref{thm:exact-support} guarantees exact reweighting by the compact LP. All steps other than the PPAD oracle are polynomial in the input size. Thus an exact rational NFCE with at most $\kappa_d$ profiles can be computed in PPAD. This meets both $\kappa_d$ and $\kappa_+$. Exact verification uses the unnormalized best-response maxima of Lemma~\ref{lem:compact-lp}. The polynomial output-length bound can be chosen to include the exact rational solutions of the compact LP. Thus both exact search problems are in PPAD, completing the proof.
\end{proof}

\begin{corollary}\label{cor:complete}
Computing an exact NFCE with support at most $\lfloor |V|^2/36\rfloor$ is PPAD-hard on instances satisfying $\kappa_d\le\lfloor |V|^2/36\rfloor$. This restriction guarantees existence. The unrestricted input domain with support size bound $\kappa_+$ instead defines a total PPAD-complete search problem.
\end{corollary}
\begin{proof}
On the restricted domain the two support size bounds $\kappa_V$ and $\kappa_+$ coincide. The restriction cannot simply be omitted for the support size bound $\kappa_V$: the simultaneous two-action matching-pennies game has $|V|=7$ and $\kappa_V=1$, but no pure Nash equilibrium, hence no NFCE supported on a single profile. The second reduction in Theorem~\ref{thm:main-hardness} lands in this domain. Totality on all inputs under $\kappa_+$ and PPAD membership follow from the exact upper bound in Theorem~\ref{thm:main-hardness}; hardness follows from its second reduction.
\end{proof}

\section{Conclusion}
We establish PPAD membership for exact NFCE computation with a rational output of quadratic support size in  perfect-recall extensive-form games. For two players quadratic support sizes give PPAD-complete search problems for exact NFCE computation. 
The questions of whether (unrestricted) any NFCE computation is PPAD-hard, and whether hardness holds under more permissive support size bounds remain open.

\bibliographystyle{plainnat}
\bibliography{biblio}

@inproceedings{anagnostides2024sparse,
  author    = {Anagnostides, Ioannis and Kalavasis, Alkis and
               Sandholm, Tuomas and Zampetakis, Manolis},
  title     = {On the Complexity of Computing Sparse Equilibria and
               Lower Bounds for No-Regret Learning in Games},
  booktitle = {Innovations in Theoretical Computer Science (ITCS)},
  year      = {2024},
  url       = {https://arxiv.org/abs/2311.14869}
}

@misc{anagnostides2026complexity,
  author        = {Anagnostides, Ioannis and Daskalakis, Constantinos and
                   Farina, Gabriele and Golowich, Noah and
                   Sandholm, Tuomas and Zhang, Brian Hu},
  title         = {On the Complexity of Correlated Equilibria Beyond
                   Normal-Form Games},
  year          = {2026},
  eprint        = {2605.17665},
  archivePrefix = {arXiv},
  note          = {Version 2, 13 November 2026},
  url           = {https://arxiv.org/abs/2605.17665v2}
}

@article{aumann1974subjectivity,
  author  = {Aumann, Robert J.},
  title   = {Subjectivity and Correlation in Randomized Strategies},
  journal = {Journal of Mathematical Economics},
  volume  = {1},
  number  = {1},
  pages   = {67--96},
  year    = {1974}
}

@misc{babichenko2014small,
  author       = {Babichenko, Yakov and Barman, Siddharth and
                  Peretz, Ron},
  title        = {Small-Support Approximate Correlated Equilibria},
  year         = {2013},
  eprint       = {1308.6025},
  archivePrefix = {arXiv},
  url          = {https://arxiv.org/abs/1308.6025}
}

@article{chen2009settling,
  author  = {Chen, Xi and Deng, Xiaotie and Teng, Shang-Hua},
  title   = {Settling the Complexity of Computing Two-Player
             {Nash} Equilibria},
  journal = {Journal of the ACM},
  volume  = {56},
  number  = {3},
  pages   = {14:1--14:57},
  year    = {2009},
  doi     = {10.1145/1516512.1516516}
}

@misc{cheval2026complexity,
  author       = {Cheval, Vincent and Horn, Florian and
                  Paul, Soumyajit and Shirmohammadi, Mahsa},
  title        = {On the Complexity of the Optimal Correlated
                  Equilibria in Extensive-Form Games},
  year         = {2026},
  eprint       = {2507.11509},
  archivePrefix = {arXiv},
  note         = {Version 2, 29 July 2026},
  url          = {https://arxiv.org/abs/2507.11509v2}
}

@book{cover2006elements,
  author    = {Cover, Thomas M. and Thomas, Joy A.},
  title     = {Elements of Information Theory},
  edition   = {2},
  publisher = {Wiley},
  year      = {2006}
}

@inproceedings{dagan2024external,
  author    = {Dagan, Yuval and Daskalakis, Constantinos and
               Fishelson, Maxwell and Golowich, Noah},
  title     = {From External to Swap Regret 2.0:
               An Efficient Reduction for Large Action Spaces},
  booktitle = {Proceedings of the ACM Symposium on
               Theory of Computing (STOC)},
  year      = {2024}
}

@inproceedings{daskalakis2025efficient,
  author    = {Daskalakis, Constantinos and Farina, Gabriele and
               Fishelson, Maxwell and Pipis, Charilaos and
               Schneider, Jon},
  title     = {Efficient Learning and Computation of Linear
               Correlated Equilibrium in General Convex Games},
  booktitle = {Proceedings of the ACM Symposium on
               Theory of Computing (STOC)},
  year      = {2025}
}

@article{etessami2021complexity,
  author  = {Etessami, Kousha},
  title   = {The Complexity of Computing a (Quasi-)Perfect
             Equilibrium for an {$n$}-Player Extensive Form Game},
  journal = {Games and Economic Behavior},
  volume  = {125},
  pages   = {107--140},
  year    = {2021},
  doi     = {10.1016/j.geb.2019.03.006}
}

@inproceedings{farina2023linear,
  author    = {Farina, Gabriele and Pipis, Charilaos},
  title     = {Polynomial-Time Linear-Swap Regret Minimization
               in Imperfect-Information Sequential Games},
  booktitle = {Advances in Neural Information Processing Systems},
  year      = {2023}
}

@inproceedings{foster2023hardness,
  author    = {Foster, Dylan J. and Golowich, Noah and
               Kakade, Sham M.},
  title     = {Hardness of Independent Learning and Sparse
               Equilibrium Computation in {Markov} Games},
  booktitle = {Proceedings of the International Conference
               on Machine Learning (ICML)},
  year      = {2023}
}

@book{gls1988geometric,
  author    = {Gr{\"o}tschel, Martin and Lov{\'a}sz, L{\'a}szl{\'o}
               and Schrijver, Alexander},
  title     = {Geometric Algorithms and Combinatorial Optimization},
  publisher = {Springer},
  year      = {1988}
}

@misc{horn2026threshold,
  author       = {Horn, Florian},
  title        = {The Threshold Problem for {NFCE} in
                  Extensive-Form Games with Chance
                  is {ExpTime}-complete},
  year         = {2026},
  howpublished = {HAL Open Science},
  note         = {Preprint, hal-05760836, version 1},
  url          = {https://hal.science/hal-05760836v1/document}
}

@inproceedings{huang2008computing,
  author    = {Huang, Wan and von Stengel, Bernhard},
  title     = {Computing an Extensive-Form Correlated Equilibrium
               in Polynomial Time},
  booktitle = {International Workshop on Internet and Network
               Economics (WINE)},
  pages     = {506--513},
  year      = {2008}
}

@article{jiang2015polynomial,
  author  = {Jiang, Albert Xin and Leyton-Brown, Kevin},
  title   = {Polynomial-Time Computation of Exact Correlated
             Equilibrium in Compact Games},
  journal = {Games and Economic Behavior},
  volume  = {91},
  pages   = {347--359},
  year    = {2015}
}

@article{kollermegiddo1992complexity,
  author  = {Koller, Daphne and Megiddo, Nimrod},
  title   = {The Complexity of Two-Person Zero-Sum Games
             in Extensive Form},
  journal = {Games and Economic Behavior},
  volume  = {4},
  number  = {4},
  pages   = {528--552},
  year    = {1992}
}

@article{kollermegiddo1996small,
  author  = {Koller, Daphne and Megiddo, Nimrod},
  title   = {Finding Mixed Strategies with Small Supports
             in Extensive Form Games},
  journal = {International Journal of Game Theory},
  volume  = {25},
  number  = {1},
  pages   = {73--92},
  year    = {1996},
  doi     = {10.1007/BF01254386}
}

@article{kollervonstengel1996efficient,
  author  = {Koller, Daphne and Megiddo, Nimrod and
             von Stengel, Bernhard},
  title   = {Efficient Computation of Equilibria for
             Extensive Two-Person Games},
  journal = {Games and Economic Behavior},
  volume  = {14},
  number  = {2},
  pages   = {247--259},
  year    = {1996}
}

@incollection{kuhn1953extensive,
  author    = {Kuhn, Harold W.},
  title     = {Extensive Games and the Problem of Information},
  booktitle = {Contributions to the Theory of Games II},
  series    = {Annals of Mathematics Studies},
  volume    = {28},
  pages     = {193--216},
  editor    = {Kuhn, Harold W. and Tucker, Albert W.},
  publisher = {Princeton University Press},
  year      = {1953}
}

@article{papadimitriou1994complexity,
  author  = {Papadimitriou, Christos H.},
  title   = {On the Complexity of the Parity Argument and
             Other Inefficient Proofs of Existence},
  journal = {Journal of Computer and System Sciences},
  volume  = {48},
  number  = {3},
  pages   = {498--532},
  year    = {1994}
}

@article{papadimitriou2008computing,
  author  = {Papadimitriou, Christos H. and Roughgarden, Tim},
  title   = {Computing Correlated Equilibria in Multi-Player Games},
  journal = {Journal of the ACM},
  volume  = {55},
  number  = {3},
  pages   = {14:1--14:29},
  year    = {2008}
}

@misc{papadimitriou2010comment,
  author = {Papadimitriou, Christos H. and Roughgarden, Tim},
  title  = {Comment on ``Computing Correlated Equilibria
            in Multi-Player Games''},
  year   = {2010},
  url    = {https://www.timroughgarden.org/papers/comment.pdf}
}

@inproceedings{peng2024fast,
  author    = {Peng, Binghui and Rubinstein, Aviad},
  title     = {Fast Swap Regret Minimization and Applications
               to Approximate Correlated Equilibria},
  booktitle = {Proceedings of the ACM Symposium on
               Theory of Computing (STOC)},
  year      = {2024},
  url       = {https://arxiv.org/abs/2310.19647}
}

@book{pinsker1964information,
  author    = {Pinsker, Mark S.},
  title     = {Information and Information Stability of
               Random Variables and Processes},
  publisher = {Holden-Day},
  year      = {1964}
}

@book{rockafellar1970convex,
  author    = {Rockafellar, R. Tyrrell},
  title     = {Convex Analysis},
  publisher = {Princeton University Press},
  year      = {1970}
}

@misc{soleymani2026support,
  author        = {Soleymani, Ashkan and Jaillet, Patrick and
                   Farina, Gabriele and Rafiey, Akbar},
  title         = {Settling the Support Complexity of
                   Correlated Equilibria},
  year          = {2026},
  eprint        = {2611.20362},
  archivePrefix = {arXiv},
  note          = {To appear in ITCS 2027},
  url           = {https://arxiv.org/abs/2611.20362}
}

@mastersthesis{stein2007thesis,
  author = {Stein, Noah},
  title  = {Characterization and Computation of Equilibria
            in Infinite Games},
  school = {Massachusetts Institute of Technology},
  year   = {2007},
  url    = {https://www.mit.edu/~nstein/documents/MastersThesis.pdf}
}

@article{vonstengel2008extensive,
  author  = {von Stengel, Bernhard and Forges, Fran{\c{c}}oise},
  title   = {Extensive-Form Correlated Equilibrium:
             Definition and Computational Complexity},
  journal = {Mathematics of Operations Research},
  volume  = {33},
  number  = {4},
  pages   = {1002--1022},
  year    = {2008}
}

@article{hart2000adaptive,
  author  = {Hart, Sergiu and Mas-Colell, Andreu},
  title   = {A Simple Adaptive Procedure Leading to Correlated Equilibrium},
  journal = {Econometrica},
  volume  = {68},
  number  = {5},
  pages   = {1127--1150},
  year    = {2000},
  doi     = {10.1111/1468-0262.00153}
}

@article{blum2007external,
  author  = {Blum, Avrim and Mansour, Yishay},
  title   = {From External to Internal Regret},
  journal = {Journal of Machine Learning Research},
  volume  = {8},
  pages   = {1307--1324},
  year    = {2007},
  url     = {https://www.jmlr.org/papers/v8/blum07a.html}
}

@article{germanolugosi2007,
  author  = {Germano, Fabrizio and Lugosi, G{\'a}bor},
  title   = {Existence of Sparsely Supported Correlated Equilibria},
  journal = {Economic Theory},
  volume  = {32},
  number  = {3},
  pages   = {575--578},
  year    = {2007}
}

@misc{anagnostides2026markov,
  author        = {Anagnostides, Ioannis and Daskalakis, Constantinos and
                   Farina, Gabriele and Golowich, Noah and
                   Sandholm, Tuomas and Zhang, Brian Hu},
  title         = {Normal-Form Correlation in {Markov} Games},
  year          = {2026},
  eprint        = {2610.03621},
  archivePrefix = {arXiv},
  url           = {https://arxiv.org/abs/2610.03621}
}

\clearpage
\appendix
\phantomsection
\addcontentsline{toc}{section}{Appendices}
\addtocontents{toc}{\protect\setcounter{tocdepth}{0}}

\section{Background on Information Theory}\label{app:information}
 For a probability distribution $p$ on a finite set $\Omega$, its entropy is $\mathsf H(p)=-\sum_{\omega\in\Omega}p(\omega)\log p(\omega)$, with $0\log0=0$. Write $\mathsf H(X)$ for the entropy of the law of $X$ and $\mathsf H(X\mid Y)=\sum_y\Pr(Y=y)\mathsf H(X\mid Y=y)$. The mutual information and relative entropy are
\begin{align*}
\mathsf I(X;Y)&=\mathsf H(X)+\mathsf H(Y)-\mathsf H(X,Y),\\
D_{\rm KL}(p\|q)&=\sum_{\omega:p(\omega)>0}p(\omega)\log\frac{p(\omega)}{q(\omega)},
\end{align*}
where relative entropy is infinite when $q$ vanishes at a positive-probability point of $p$. We use the standard chain rule, nonnegativity of conditional mutual information, and the identity $\mathsf I(X;Y)=D_{\rm KL}(\mathcal L(X,Y)\|\mathcal L(X)\otimes\mathcal L(Y))$; see \citep{cover2006elements}. In particular, entropy on a set of size $k$ is at most $\log k$, and applying a deterministic function to either variable cannot increase their mutual information. The latter is the data-processing inequality.

For finite distributions $p,q$ on the same set, total variation is $\operatorname{TV}(p,q)=\frac12\sum_\omega|p(\omega)-q(\omega)|$.
\begin{lemma}[Pinsker's inequality~[\cite{pinsker1964information}]]\label{lem:pinsker}
For finite distributions $p,q$,
\[
\operatorname{TV}(p,q)\le\sqrt{D_{\rm KL}(p\|q)/2}.
\]
Moreover, for every function $f:\Omega\to[0,1]$,
$|\mathbb E_p f-\mathbb E_q f|\le\operatorname{TV}(p,q)$.
\end{lemma}

\section{Uniform rational LP gap}\label{app:gap}
\begin{proof}[Proof of Lemma~\ref{lem:uniform-gap}]
As in Section~\ref{sec:choosing-epsilon}, terminal payoffs are normalized to $[0,1]$ and $K=1+\sum_i d_i(d_i+1)\le1+N(N-1)$.
For each leaf $z$, the denominator of $p_c(z)u_i(z)$ divides $B$, because a root--leaf path uses each chance edge at most once. By \eqref{eq:payoff-leaves}, all pure-profile payoffs lie in $B^{-1}\mathbb Z\cap[0,1]$ and all pure deviation differences lie in $B^{-1}\mathbb Z\cap[-1,1]$. Let $S=\{\tau^1,\ldots,\tau^k\}$ with $k\le K$, and set $d:=k+1$.
The LP \eqref{eq:conceptual-lp} has $d$ variables, namely
$(\mu(\tau^1),\ldots,\mu(\tau^k),t)$. Its feasible region is nonempty because any
probability vector together with $t=1$ is feasible, and it is bounded
because $\mu(\tau)\ge0$, $\sum_{\tau\in S}\mu(\tau)=1$, and $0\le t\le 1$.
Hence an optimal vertex $x^*$ exists.

Choose $d$ linearly independent active constraints at $x^*$ and write
them as
\[
Mx^*=r,
\]
where $M\in\mathbb Q^{d\times d}$ is nonsingular. Every entry of $M$
and every entry of $r$ is a rational number with denominator dividing
$B$, and every entry of $M$ has absolute value at most $1$.
Multiplying by $B$ gives an equivalent integer system
\[
Ax^*=c,
\]
where $A\in\mathbb Z^{d\times d}$ is nonsingular,
$c\in\mathbb Z^d$, and every entry of $A$ has absolute value at most
$B$.

If $t^*>0$, then Cramer's rule gives
\[
t^*=\frac{\det(A_t)}{\det(A)},
\]
where $A_t$ is obtained from $A$ by replacing the column of $t$ by
$c$. Since $A_t$ is an integer matrix and $t^*>0$, we have
$|\det(A_t)|\ge 1$. Therefore
\[
t^*\ge \frac{1}{|\det(A)|}.
\]
Each row of $A$ has Euclidean norm at most $\sqrt d\,B$, so
Hadamard's determinant inequality yields
\[
|\det(A)|\le (\sqrt d\,B)^d\le (dB)^d.
\]
It follows that
\[
t^*>0
\quad\Longrightarrow\quad
t^*\ge (dB)^{-d}.
\]
Since $d\le K+1$ and $B\ge 1$,
\[
(dB)^{-d}\ge ((K+1)B)^{-(K+1)}=\eta.
\]
Thus $t^*(S)=0$ or $t^*(S)\ge \eta$.

Finally,
\[
\log_2(1/\eta)=(K+1)\bigl(\log_2(K+1)+\log_2 B\bigr),
\]
which is polynomial in the input encoding length because $K$ and
$\log B$ are polynomial in the explicit game size.
\end{proof}

\end{document}